\documentclass[preprint]{sigplanconf}

\usepackage{bm}
\usepackage{pgfplots}
\usepackage{enumerate,paralist}
\usepackage{pbox}

\usepackage{appendix}
\usepackage{comment}
\usepackage{parskip}
\usepackage{multirow}
\usepackage{url}

\usepackage[ruled, linesnumbered]{algorithm2e}
\usepackage{tikz,pgffor}
\usetikzlibrary{arrows}
\usetikzlibrary{shapes}
\usetikzlibrary{calc}
\usetikzlibrary{automata}
\usetikzlibrary{positioning}

\tikzstyle{ang}=[regular polygon, regular polygon sides = 3,draw,inner sep=0pt,minimum size=6mm, yshift = -0.75 mm]
\tikzstyle{dem}=[shape=diamond,draw,inner sep=0pt,minimum size=6mm]
\tikzstyle{ran}=[shape=circle,draw,inner sep=0pt,minimum size=5mm]
\tikzstyle{det}=[shape=rectangle,draw,inner sep=0pt,minimum size=5mm]
\tikzstyle{tran}=[draw,->,>=stealth, rounded corners]

\usepackage{calc}
\usepackage{fp}

\usepackage{lmodern}
\usepackage{amsmath}
\usepackage{amssymb, amsthm}
\usepackage{listings}
\usepackage{mathtools}
\usepackage{times}

\usepackage{graphicx}

\theoremstyle{plain}
\newtheorem{lemma}{Lemma}

\newtheorem{proposition}{Proposition}
\newtheorem{definition}{Definition}
\newtheorem{notations}{Notations}
\newtheorem{corollary}{Corollary}
\newtheorem{theorem}{Theorem}
\newtheorem{example}{Example}
\newtheorem{remark}{Remark}

\newcommand{\expv}{\mathbb{E}}
\newcommand{\Eval}{\mathsf{ET}}
\newcommand{\Thr}{\mathsf{Thr}}

\lstdefinelanguage{affprob}
{
morekeywords={angel,demon, prob, if, then, else, fi, while, do, od, true, false, and, or, skip},
sensitive = false
}

\DeclareMathAlphabet{\mathpzc}{OT1}{pzc}{m}{it}

\renewcommand{\vec}[1]{\mathbf{#1}}

\newcommand{\Nats}{\mathbb{N}}

\newcommand{\E}{\ensuremath{{\rm \mathbb E}}}

\newcommand{\eps}{\epsilon}

\newcommand{\wh}{\widehat}

\newcommand{\pvars}{X}
\newcommand{\rvars}{R}
\newcommand{\locs}{\mathit{L}}

\newcommand{\loc}{\ell}
\newcommand{\initloc}{\loc_0}
\newcommand{\initval}{\mathbf{x}_0}
\newcommand{\tran}{\tau}
\newcommand{\lit}{\xi}
\newcommand{\satpenalty}{\mathit{Pen}}

\newcommand{\mart}{\eta}
\renewcommand{\E}{\mathbb{E}}

\newcommand{\Rset}{\mathbb{R}}
\newcommand{\Nset}{\mathbb{N}}
\newcommand{\Zset}{\mathbb{Z}}

\newcommand{\lin}{\mathit{in}}
\newcommand{\lout}{\mathit{out}}

\newcommand{\transitions}{\mapsto}
\newcommand{\probdist}{\mathit{Pr}}
\newcommand{\guards}{G}

\newcommand{\prob}{\mathit{Pr}}
\newcommand{\assgn}[2]{[#1/#2]}
\newcommand{\id}{\mathit{id}}
\newcommand{\probm}{\mathbb{P}}
\newcommand{\inv}{I}

\newcommand{\APP}{{\sc App}}

\newcommand{\LRAPP}{{\sc LRApp}}

\newcommand{\NP}{\ensuremath{\mathsf{NP}}}

\newcommand{\PSPACE}{\ensuremath{\mathsf{PSPACE}}}
\newcommand{\TWOEXPTIME}{\ensuremath{\mathsf{2EXPTIME}}}

\newcommand{\Op}{\mathsf{Op}}
\newcommand{\Operator}{\mathsf{SetOp}}
\newcommand{\LRSMSynth}{{\sc LRSMSynth}}
\newcommand{\Basic}{\mathsf{Half}}
\newcommand{\OneStep}{\mathsf{PreExp}}
\newcommand{\NS}[1]{\mathsf{NS}(#1)}
\newcommand{\UB}{\mathsf{UB}}

\newcommand{\costvar}{W}
\newcommand{\stepcost}{cost}

\newcommand{\unfold}[2]{\mathit{Unf}(#1,#2)}

\newcommand{\term}{Term}
\newcommand{\martcor}[1]{\mathit{cmp(#1)}}

\title{Algorithmic Analysis of Qualitative and Quantitative Termination Problems for Affine Probabilistic Programs\thanks{The research was partly supported by Austrian Science Fund (FWF) Grant No P23499- N23, FWF NFN Grant No S11407-N23 (RiSE/SHiNE), and ERC Start grant (279307: Graph Games). The research leading to these results has received funding from the People Programme (Marie
Curie Actions) of the European Union's Seventh Framework Programme (FP7/2007-2013) under
REA grant agreement No [291734].}}

\authorinfo{Krishnendu Chatterjee}{IST Austria}{Krishnendu.Chatterjee@ist.ac.at}
\authorinfo{Hongfei Fu\thanks{Supported by the Natural Science Foundation of China (NSFC) under Grant No. 61532019 .}}{IST Austria \\State Key Laboratory of Computer Science, Institute of Software, Chinese Academy of Sciences}{hongfei.fu@ist.ac.at}
\authorinfo{Petr Novotn\'{y}}{IST Austria}{petr.novotny@ist.ac.at}
\authorinfo{Rouzbeh Hasheminezhad}{Sharif University of Technology}{hasheminezhad@ce.sharif.edu}

\begin{document}
\maketitle
\begin{abstract}
In this paper, we consider termination of probabilistic programs with real-valued 
variables.
The questions concerned are:
\begin{compactenum}
\item qualitative ones that ask (i) whether the program terminates with probability $1$ (almost-sure termination) and (ii) whether the expected termination time is finite (finite termination); 
\item quantitative ones that ask (i) to approximate the expected termination time (expectation problem) and (ii) to compute a bound $B$ such that the probability to terminate after $B$ steps decreases exponentially (concentration problem).
\end{compactenum}
To solve these questions, we utilize the notion of ranking supermartingales which is a powerful approach for proving termination of probabilistic programs.
In detail, we focus on algorithmic synthesis of linear ranking-supermartingales over affine probabilistic programs (\APP's) with both angelic and demonic non-determinism. 
An important subclass of \APP's is \LRAPP~which is defined as the class of all \APP's over which a linear ranking-supermartingale exists.

Our main contributions are as follows.
Firstly, we show that the membership problem of \LRAPP~(i) can be decided in polynomial time for \APP's with at most demonic non-determinism, and (ii) is $\NP$-hard and in $\PSPACE$ for 
\APP's with angelic non-determinism; 
moreover, the $\NP$-hardness result holds already for \APP's without probability and demonic non-determinism. 
Secondly, we show that the concentration problem over \LRAPP~can be solved in the same complexity as for the membership problem of \LRAPP.
Finally, we show that the expectation problem over \LRAPP~can be solved in $\TWOEXPTIME$ and is $\PSPACE$-hard even for \APP's without probability and non-determinism (i.e., deterministic programs). 
Our experimental results demonstrate the effectiveness of our approach to answer
the qualitative and quantitative questions over \APP's with at most demonic non-determinism. 
\end{abstract}

\section{Introduction}\label{sec:introduction}

\paragraph*{Probabilistic Programs}
Probabilistic programs extend the classical imperative programs
with \emph{random-value generators} that produce random values
according to some desired probability distribution.
They provide a rich framework to model a
wide variety of applications ranging from randomized
algorithms~\cite{RandBook,RandBook2}, to stochastic network
protocols~\cite{BaierBook,prism},
to robot planning~\cite{KGFP09,kaelbling1998planning}, just to mention a few.
The formal analysis of probabilistic systems in general and
probabilistic programs in particular has received a lot of attention
in different areas, such as probability theory and
statistics~\cite{Durrett,Howard,Kemeny,Rabin63,PazBook},
formal methods~\cite{BaierBook,prism},
artificial intelligence~\cite{LearningSurvey,kaelbling1998planning},
and programming languages~\cite{SriramCAV,HolgerPOPL,SumitPLDI,EGK12}.

\paragraph*{Qualitative and Quantitative Termination
Questions}
The most basic, yet important, notion of liveness for programs
is \emph{termination}.
For non-probabilistic programs, proving termination is equivalent
to synthesizing \emph{ranking functions}~\cite{rwfloyd1967programs}, and many different
approaches exist for synthesis of ranking functions over
non-probabilistic programs~\cite{DBLP:conf/cav/BradleyMS05,DBLP:conf/tacas/ColonS01,DBLP:conf/vmcai/PodelskiR04,DBLP:conf/pods/SohnG91}.
While a ranking function guarantees termination of a non-probabilistic program with certainty in a finite number of steps, there are many natural extensions of the termination
problem in the presence of probability.
In general, we can classify the termination questions over probabilistic programs as \emph{qualitative} and \emph{quantitative} ones.
The relevant questions studied in this paper are illustrated as follows.
\begin{compactenum}
\item \emph{Qualitative Questions.}
The most basic qualitative question is on \emph{almost-sure termination}
which asks whether a program terminates with probability~1.
Another fundamental question is about \emph{finite termination} (aka positive almost-sure
termination~\cite{HolgerPOPL,BG05}) which asks whether the expected
termination time is finite.
Note that finite expected termination time implies almost-sure termination,
whereas the converse does not hold in general.
\item \emph{Quantitative Questions.}
We consider two quantitative questions, namely \emph{expectation} and \emph{concentration} questions.
The expectation question asks to approximate the expected termination time of a probabilistic program (within some additive or relative error), provided that the expected termination time is finite.
The concentration problem asks to compute a bound $B$ such that the
probability that the termination time is below $B$ is concentrated,
or in other words, the probability that the termination time exceeds the
bound $B$ decreases exponentially.
\end{compactenum}
Besides, we would like to note that there exist other quantitative questions such as \emph{bounded-termination} question which asks to approximate the probability to terminate after a given number of steps (cf.~\cite{DBLP:conf/sas/Monniaux01} etc.).

\paragraph*{Non-determinism in Probabilistic Programs}
Along with probability, another fundamental aspect in modelling is
\emph{non-determinism}.
In programs, there can be two types of non-determinism:
(i) \emph{demonic} non-determinism that is adversarial (e.g., to be resolved to ensure non-termination or to increase the expected termination time, etc.) and
(ii) \emph{angelic} non-determinism that is favourable
(e.g., to be resolved to ensure termination or to decrease the
expected termination time, etc.).
The demonic non-determinism is necessary in many cases, and a classic
example is abstraction:
for efficient static analysis of large programs, it is infeasible to track
all variables of the program;
the key technique in such cases is abstraction of variables, where
certain variables are not considered for the analysis and they are instead assumed to induce a worst-case behaviour, which exactly corresponds to demonic
non-determinism.
On the other hand, angelic non-determinism is relevant in synthesis.
In program sketching (or programs with holes as studied extensively
in~\cite{Sketching}), certain expressions can be synthesized which helps in
termination, and this corresponds to resolving non-determinism in an
angelic way.
The consideration of the two types of non-determinism gives the following
classes:
\begin{compactenum}
\item probabilistic programs without non-determinism;
\item probabilistic programs with at most demonic non-determinism;
\item probabilistic programs with at most angelic non-determinism;
\item probabilistic programs with both angelic and demonic non-determinism.
\end{compactenum}

\paragraph*{Previous Results}
We discuss the relevant previous results for termination analysis of probabilistic programs.
\begin{compactitem}
\item \emph{Discrete Probabilistic Choices.}
In~\cite{MM04,MM05}, McIver and Morgan presented quantitative invariants
to establish termination, which works for probabilistic programs with
non-determinism, but restricted only to discrete probabilistic choices.

\item \emph{Infinite Probabilistic Choices without Non-determinism.}
On one hand, the approach of~\cite{MM04,MM05} was extended in~\cite{SriramCAV} to \emph{ranking  supermartingales} resulting in a sound (but not complete) approach to prove
almost-sure termination for infinite-state probabilistic programs with
integer- and real-valued random variables drawn from distributions including
uniform, Gaussian, and Poison;
the approach was only for probabilistic programs without non-determinism.
On the other hand, Bournez and Garnier~\cite{BG05} related the termination
of probabilistic programs without non-determinism to \emph{Lyapunov ranking functions}.
For probabilistic programs with countable state-space and without
non-determinism, the Lyapunov ranking functions provide a sound and complete
method for proving finite termination~\cite{BG05,Foster53}.
Another relevant approach~\cite{DBLP:conf/sas/Monniaux01} is to explore the exponential decrease of probabilities upon bounded-termination through abstract interpretation~\cite{DBLP:conf/popl/CousotC77}, resulting in a sound method for proving almost-sure termination.

\item \emph{Infinite Probabilistic Choices with Non-determinism.}
The situation changes significantly in the presence of non-determinism.
The Lyapunov-ranking-function method as well as the ranking-supermartingale method
are sound but not complete in the presence of non-determinism for finite termination~\cite{HolgerPOPL}.
However, for a subclass of probabilistic programs with at most demonic non-determinism, a sound and complete characterization for finite termination through ranking-supermartingale is obtained in~\cite{HolgerPOPL}.
\end{compactitem}

\paragraph*{Our Focus}
We focus on ranking-supermartingale based algorithmic study for qualitative and
quantitative questions on termination analysis of probabilistic programs with non-determinism.
In view of the existing results, there are at least three important classes of
open questions, namely (i) efficient algorithms, (ii) quantitative questions and (iii) complexity in presence of two different types of non-determinism.
Firstly, while~\cite{HolgerPOPL} presents a fundamental result
on ranking supermartingales over probabilistic programs with non-determinism, the generality of the result makes it difficult to obtain efficient algorithms;
hence an important open question that has not been addressed before is
whether efficient algorithmic approaches can be developed for synthesizing ranking supermartingales of simple form over probabilistic programs with
non-determinism.
The second class of open questions asks whether ranking supermartingales can be used to answer quantitative questions, which have not been tackled at all to our knowledge.
Finally, no previous work considers complexity to analyze probabilistic programs with both the two types of non-determinism (as required for the synthesis problem with abstraction).

\begin{table*}[t]
\begin{center}
{\small
\begin{tabular}{|c|c|c|c|c|}
\hline
\hline
Questions /\/ Models &  \;Prob prog without nondet\;    & \;Prob prog with demonic nondet\;        &  \;Prob prog with angelic nondet\;    & \;Prob prog with both nondet\; \\
\hline
Qualitative & & & $\NP$-hard & $\NP$-hard \\
(Almost-sure,  & PTIME & PTIME & &  \\
Finite termination)  & & & $\PSPACE$ & $\PSPACE$ \\
                     & & & (QCQP)   & (QCQP) \\
\hline

Quantitative & $\PSPACE$-hard & $\PSPACE$-hard & $\PSPACE$-hard & $\PSPACE$-hard \\
(Expectation)  & $\TWOEXPTIME$ & $\TWOEXPTIME$ & $\TWOEXPTIME^\dagger$ & $\TWOEXPTIME^\dagger$ \\
\hline
\hline
\end{tabular}
}
\end{center}
\caption{Computational complexity of qualitative and quantitative questions for termination of probabilistic programs in \LRAPP, where the complexity of
quantitative questions is for bounded \LRAPP s with discrete probability choices. Results marked by $\dagger$ were obtained under additional assumptions. \label{tab:complexity}}
\vspace{-1em}
\end{table*}

\paragraph*{Our Contributions}
In this paper, we consider a subclass of probabilistic programs called \emph{affine probabilistic programs (\APP's)} which involve both demonic and angelic non-determinism.
In general, an \APP~is a probabilistic program whose all arithmetic expressions are linear.
Our goal is to analyse the simplest class of ranking supermartingales over \APP's,
namely, \emph{linear ranking supermartingales}.
We denote by \LRAPP~the set of all \APP's that admit a linear ranking supermartingale.
Our main contributions are as follows:
\begin{compactenum}
\item \emph{Qualitative Questions.} Our results are as follows.

\smallskip\noindent{\em Algorithm.}
We present an algorithm for probabilistic programs with both
angelic and demonic non-determinism that decides whether a given instance
of an \APP\ belongs to \LRAPP\ (i.e., whether a linear ranking supermartingale
exists), and if yes, then synthesize a linear ranking supermartingale (for proving almost-sure termination).
We also show that almost-sure termination coincides with finite termination over \LRAPP.
Our result generalizes the one~\cite{SriramCAV} for probabilistic programs
without non-determinism to probabilistic programs with both the two types of
non-determinism. 
Moreover, in~\cite{SriramCAV} even for affine probabilistic programs without non-determinism, possible quadratic constraints may be constructed; in contrast, we show that for affine probabilistic programs with at most demonic non-determinism, a set of linear constraints suffice, leading to polynomial-time decidability (cf. Remark~\ref{rmk:linearconstraintsptime}). 

\smallskip\noindent{\em Complexity.}
We establish a number of complexity results as well.
For programs in \LRAPP~ with at most demonic non-determinism our algorithm runs in polynomial time
by reduction to solving a set of linear constraints.
In contrast, we show that for probabilistic programs in \APP's with only angelic
non-determinism even deciding whether a given instance belongs
to \LRAPP\ is $\NP$-hard.
In fact our hardness proof applies even in the case when there are
no probabilities but only angelic non-determinism.
Finally, for \APP's with two types of non-determinism (which is $\NP$-hard as the
special case with only angelic non-determinism is $\NP$-hard) our algorithm reduces
to quadratic constraint solving.
The problem of quadratic constraint solving is also $\NP$-hard and can be solved in
$\PSPACE$; we note that developing practical approaches to quadratic constraint solving
(such as using semidefinite relaxation) is an active research area~\cite{DBLP:books/el/RV01/BockmayrW01}.

\item \emph{Quantitative Questions.}
We present three types of results. To the best of our knowledge, we present the
first complexity results (summarized in Table~\ref{tab:complexity}) for
quantitative questions.
First, we show that the expected termination time is irrational in general for programs
in \LRAPP.
Hence we focus on the approximation questions.
For concentration results to be applicable, we consider the class {\em bounded \LRAPP}
which consists of programs that admit a linear ranking supermartingale with bounded difference. Our results are as follows.

\smallskip\noindent{\em Hardness Result.}
We show that the expectation problem is
$\PSPACE$-hard even for deterministic programs in bounded \LRAPP.

\smallskip\noindent{\em Concentration Result on Termination Time.}
We present the first concentration result on termination time through linear ranking-supermartingales over probabilistic programs in bounded \LRAPP.
We show that by solving a variant version of the problem for the
qualitative questions, 
we can obtain a bound $B$ such that the probability that the termination time
exceeds $n \geq B$ decreases exponentially in $n$.
Moreover, the bound $B$ computed is at most exponential.
As a consequence,
unfolding a program upto $O(B)$ steps and approximating the expected termination time
explicitly upto $O(B)$ steps, imply approximability (in $\TWOEXPTIME$)
for the expectation problem.

\smallskip\noindent {\em Finer Concentration Inequalities.}
Finally, in analysis of supermartingales for probabilistic programs only  Azuma's
inequality~\cite{Azuma1967inequality} has been proposed in the literature~\cite{SriramCAV}.
We show how to obtain much finer concentration inequalities using
Hoeffding's inequality~\cite{Hoeffding1963inequality,ColinMcDiarmid1998concentration}
(for all programs in bounded \LRAPP)
and Bernstein's inequalities~\cite{Berstein1962inequality,ColinMcDiarmid1998concentration} (for incremental programs in \LRAPP, where all updates are increments/decrements by some affine expression over random variables).
Bernstein's inequality is based on the deep mathematical results
in measure theory on spin glasses~\cite{Berstein1962inequality}, and we show how they can be used for analysis of probabilistic programs.

\smallskip\noindent {\em Experimental Results.}
We show the effectiveness of our approach to answer qualitative and concentration
questions on several classical problems, such as random walk in one dimension,
adversarial random walk in one dimension and two dimensions (that involves both
probability and demonic non-determinism).

\end{compactenum}

Note that the most restricted class we consider is bounded \LRAPP,
but we show that several classical problems, such as random walks in one
dimension, queuing processes, belong to bounded \LRAPP, for which
our results provide a practical approach.

\section{Preliminaries}

\subsection{Basic Notations}
For a set $A$ we denote by $|A|$ the cardinality of $A$. We denote by $\Nset$, $\Nset_0$, $\Zset$, and $\Rset$ the sets of all positive integers, non-negative integers, integers, and real numbers, respectively. We use boldface notation for vectors, e.g. $\vec{x}$, $\vec{y}$, etc, and we denote an $i$-th component of a vector $\vec{x}$ by $\vec{x}[i]$.

An \emph{affine expression} is an expression of the form $d+\sum_{i=1}^{n}a_i x_i$, where $x_1,\dots,x_n$ are variables and $d,a_1,\dots,a_n$ are real-valued constants.
Following the terminology of~\cite{DBLP:conf/sas/KatoenMMM10} we fix the following nomenclature:
\begin{compactitem}
\item {\em Linear Constraint.} A \emph{linear constraint} is a formula of the form $\psi$ or $\neg\psi$, where  $\psi$ is a non-strict inequality
between affine expressions.
\item {\em Linear Assertion.} A \emph{linear assertion} is a finite conjunction of linear constraints.
\item {\em Propositionally Linear Predicate.}
A  \emph{propositionally linear predicate} is a finite disjunction of linear assertions.
\end{compactitem}
In this paper, we deem any linear assertion equivalently as a polyhedron defined by the linear assertion (i.e., the set of points satisfying the assertion).
It will be always clear from the context whether a linear assertion is deemed as a logical formula or as a polyhedron.

\subsection{Syntax of Affine Probabilistic Programs}

In this subsection, we illustrate the syntax of programs that we study.
We refer to this class of programs as \emph{affine probabilistic programs} since it involves solely affine expressions.

Let $\mathcal{X}$ and $\mathcal{R}$ be countable collections of \emph{program} and \emph{random} variables, respectively. The abstract syntax of affine probabilistic programs (\APP s)
is given by the grammar in Figure~\ref{fig:syntax}, where
the expressions $\langle \mathit{pvar}\rangle$ and $\langle \mathit{rvar}\rangle$  range over $\mathcal{X}$ and $\mathcal{R}$, respectively.
The grammar is such that $\langle \mathit{expr} \rangle$ and $\langle \mathit{rexpr} \rangle$ may evaluate to an arbitrary affine expression over the program variables, and the program and random variables, respectively (note that random variables can only be used in the RHS of an assignment). Next, $\langle \mathit{bexpr}\rangle$ may evaluate to an arbitrary propositionally linear predicate.

The guard of each if-then-else statement is either a keyword \textbf{angel} (intuitively, this means that the fork is non-deterministic and the non-determinism is resolved angelically; see also the definition of semantics below), a keyword \textbf{demon} (demonic resolution of non-determinism), keyword \textbf{prob}($p$), where $p\in [0,1]$ is a number given in decimal representation (represents probabilistic choice, where  the if-branch is executed with probability $p$ and the then-branch with probability $1-p$), or the guard is a propositionally linear predicate, in which case the statement represents a standard deterministic conditional branching.

\begin{figure}
\begin{align*}
\langle \mathit{stmt}\rangle &::= \,\langle\mathit{pvar}\rangle \,\text{'$:=$'}\, \langle\mathit{rexpr} \rangle \\
&\mid   \text{'\textbf{if}'} \, \langle\mathit{ndbexpr}\rangle\,\text{'\textbf{then}'} \, \langle \mathit{stmt}\rangle \, \text{'\textbf{else}'} \, \langle \mathit{stmt}\rangle \,\text{'\textbf{fi}'}
\\
&\mid  \text{'\textbf{while}'}\, \langle\mathit{bexpr}\rangle \, \text{'\textbf{do}'} \, \langle \mathit{stmt}\rangle \, \text{'\textbf{od}'}
\\
&\mid \langle\mathit{stmt}\rangle \, \text{';'} \, \langle \mathit{stmt}\rangle \mid \text{'\textbf{skip}'}
\\
\vspace{0.5\baselineskip}
\vspace{0.5\baselineskip}
\langle\mathit{expr} \rangle &::= \langle \mathit{constant} \rangle \mid \langle\mathit{pvar}\rangle
\mid \langle \mathit{constant} \rangle \,\text{'$*$'} \, \langle\mathit{pvar}\rangle
\\
&\mid \langle\mathit{expr} \rangle\, \text{'$+$'} \,\langle\mathit{expr} \rangle \mid \langle\mathit{expr} \rangle\, \text{'$-$'} \,\langle\mathit{expr} \rangle
\\
\vspace{0.5\baselineskip}
\langle\mathit{rexpr} \rangle &::= \langle\mathit{expr} \rangle \mid  \langle\mathit{rvar}\rangle \mid \langle\mathit{pvar}\rangle \mid\langle\mathit{constant} \rangle \,\text{'$*$'} \, \langle\mathit{rvar}\rangle \\&\mid\langle\mathit{constant} \rangle \,\text{'$*$'} \, \langle\mathit{pvar}\rangle
\mid \langle\mathit{rexpr} \rangle\, \text{'$+$'} \,\langle\mathit{rexpr} \rangle \\ &\mid \langle\mathit{rexpr} \rangle\, \text{'$-$'} \,\langle\mathit{rexpr} \rangle
\\
\vspace{0.5\baselineskip}
\langle \mathit{bexpr}\rangle &:=  \langle \mathit{affexpr} \rangle \mid \langle \mathit{affexpr} \rangle \, \text{'\textbf{or}'} \, \langle\mathit{bexpr}\rangle
\vspace{0.5\baselineskip}
\\
\langle\mathit{affexpr} \rangle &::=  \langle\mathit{literal} \rangle\mid \langle\mathit{literal} \rangle\, \text{'\textbf{and}'} \,\langle\mathit{affexpr} \rangle
\\
\langle\mathit{literal} \rangle &::= \langle\mathit{expr} \rangle\, \text{'$\leq$'} \,\langle\mathit{expr} \rangle \mid \langle\mathit{expr} \rangle\, \text{'$\geq$'} \,\langle\mathit{expr} \rangle
\\
&\mid \neg \langle \mathit{literal} \rangle
\\
\langle\mathit{ndbexpr} \rangle &::= \text{'\textbf{angel}'} \mid \text{'\textbf{demon}'}\mid \text{'\textbf{prob($p$)}'} \mid \langle\mathit{bexpr} \rangle
\end{align*}
\caption{Syntax of affine probabilistic programs (\APP 's).}
\label{fig:syntax}
\end{figure}



\begin{example}\label{ex:prog}
We present an example of an affine probabilistic program shown in Figure~\ref{ex:prob}.
The program variable is $x$, and there is a while loop, where given a probabilistic
choice one of two statement blocks $Q_1$ or $Q_2$ is executed.
The block $Q_1$ (resp. $Q_2$) is executed if the probabilistic choice is at least
$0.6$ (resp. less than $0.4$).
The statement block $Q_1$ (resp., $Q_2$) is an angelic (resp. demonic) conditional
statement to either increment or decrement $x$.
\lstset{language=affprob}
\lstset{tabsize=3}
\newsavebox{\affproblist}
\begin{lrbox}{\affproblist}
\begin{lstlisting}[mathescape]
$x:=0$;
while $x \geq 0$ do
	if prob(0.6) then
		if angel then
			$x:=x+1$
		else
			$x:=x-1$
		fi
	else
		if demon then
			$x:=x+1$
		else
			$x:=x-1$
		fi
	fi
od
\end{lstlisting}
\end{lrbox}
\begin{figure}[t]
\usebox{\affproblist}
\caption{An example of a probabilistic program}\label{ex:prob}
\end{figure}
\end{example}

\subsection{Semantics of Affine Probabilistic Programs}


We now formally define the semantics of \APP's.
In order to do this, we first recall some fundamental concepts from probability theory.

\paragraph*{Basics of Probability Theory}
The crucial notion is of the probability space. A probability space is a triple $(\Omega,\mathcal{F},\probm)$, where $\Omega$ is a non-empty set (so called \emph{sample space}), $\mathcal{F}$ is a \emph{sigma-algebra} over $\Omega$, i.e. a collection of subsets of $\Omega$ that contains the empty set $\emptyset$, and that is closed under complementation and countable unions, and $\probm$ is a \emph{probability measure} on $\mathcal{F}$, i.e., a function $\probm\colon \mathcal{F}\rightarrow[0,1]$ such that
\begin{compactitem}
\item $\probm(\emptyset)=0$,
\item for all $A\in \mathcal{F}$ it holds $\probm(\Omega\smallsetminus A)=1-\probm(A)$, and
\item for all pairwise disjoint countable set sequences $A_1,A_2,\dots \in \mathcal{F}$ (i.e., $A_i \cap A_j = \emptyset$ for all $i\neq j$)
we have $\sum_{i=1}^{\infty}\probm(A_i)=\probm(\bigcup_{i=1}^{\infty} A_i)$.
\end{compactitem}

A \emph{random variable} in a probability space $(\Omega,\mathcal{F},\probm)$ is an $\mathcal{F}$-measurable function $X\colon \Omega \rightarrow \Rset \cup \{\infty\}$, i.e.,
a function such that for every $x\in \Rset \cup \{ \infty\}$ the set $\{\omega\in \Omega\mid X(\omega)\leq x\}$ belongs to $\mathcal{F}$.
We denote by $\expv(X)$ the \emph{expected value} of a random variable $X$, i.e. the Lebesgue integral of $X$ with respect to the probability measure $\probm$.
The precise definition of the Lebesgue integral of $X$ is somewhat technical and we omit it here, see,
e.g., ~\cite[Chapter 4]{Rosenthal:book}, or~\cite[Chapter 5]{Billingsley:book} for a formal definition.
A \emph{filtration} of a probability space $(\Omega,\mathcal{F},\probm)$ is a sequence $\{\mathcal{F}_i \}_{i=0}^{\infty}$ of $\sigma$-algebras over $\Omega$ such that $\mathcal{F}_0 \subseteq \mathcal{F}_1 \subseteq \cdots \subseteq \mathcal{F}_n \subseteq \cdots \subseteq \mathcal{F}$.

\paragraph*{Stochastic Game Structures}
There are several ways in which one can express the semantics of \APP's with (angelic and demonic) non-determinism~\cite{SriramCAV,HolgerPOPL}.
In this paper we take an operational approach, viewing our programs as 2-player stochastic games,
where one-player represents the angelic non-determinism, and the other player (the opponent) the demonic non-determinism.

\begin{definition}
\label{def:stochgame}
A \emph{stochastic game structure (SGS)} is a tuple $\mathcal{G}=(\locs,(\pvars,\rvars),\ell_0,\vec{x}_0,\transitions,\probdist,\guards)$, where
\begin{compactitem}
\item $\locs$ is a finite set of \emph{locations} partitioned into four pairwise disjoint subsets  $\locs_A$, $\locs_D$, $\locs_P$, and $\locs_S$ of angelic, demonic, probabilistic, and standard (deterministic) locations;
\item $\pvars$ and $\rvars$ are finite disjoint sets of real-valued \emph{program} and \emph{random variables}, respectively. We denote by $\mathcal{D}$ the joint distribution of variables in $\rvars$;
\item $\ell_0$ is an initial location and $\vec{x}_0$ is an initial valuation of program variables;
\item $\transitions$ is a transition relation, whose every member is a tuple of the form $(\ell,f,\ell')$, where $\ell$ and $\ell'$ are source and target program locations, respectively, and $f\colon \Rset^{|\pvars\cup\rvars|}\rightarrow \Rset^{|\pvars|}$ is an \emph{update function};
\item $\probdist=\{\prob_{\ell}\}_{\ell \in \locs_P}$ is a collection of
probability distributions, where each $\prob_{\ell}$ is a discrete probability
distribution on the set of all transitions outgoing from~$\ell$.
\item $\guards$ is a function assigning a propositionally linear predicates (\emph{guards}) to each transition outgoing from deterministic locations.
\end{compactitem}

We stipulate that each location has at least one outgoing transition.
Moreover, for every deterministic location $\ell$ we assume the following: if $\tau_1,\dots,\tau_k$ are all transitions outgoing from $\ell$, then $G(\tau_1) \vee \dots \vee G(\tau_k) \equiv \mathit{true}$ and $G(\tau_i) \wedge G(\tau_j) \equiv \mathit{false}$ for each $1\leq i < j \leq k$.
And we assume that each coordinate of $\mathcal{D}$ represents an integrable random variable (i.e., the expected value of the absolute value of the random variable exists).
\end{definition}

For notational convenience we assume that the sets $\pvars$ and $\rvars$ are endowed with some fixed linear ordering, which allows us to write $\pvars = \{x_1,x_2,\dots,x_{|\pvars|}\}$ and $\rvars = \{r_1,r_2,\dots,r_{|\rvars|}\}$. Every update function $f$ in a stochastic game can then be viewed as a tuple $(f_1,\dots,f_{|\pvars|})$, where each $f_i$ is of type $\Rset^{|\pvars\cup\rvars|}\rightarrow\Rset$. We denote by $\mathbf{x}=(\vec{x}[i])_{i=1}^{|\pvars|}$ and $\mathbf{r}=(\vec{r}[i])_{i=1}^{|\rvars|}$ the vectors of concrete valuations of program and random variables, respectively.
In particular, we assume that each component of $\vec{r}$ lies within the range of the corresponding random variable. 
We use the following succinct notation for special update functions: by $\id$ we denote a function which does not change the program variables at all, i.e. for every $1\leq i \leq |\pvars|$ we have $f_i(\vec{x},\vec{r})= \vec{x}[i]$. For a function $g$ over the program and random variables we denote by $\assgn{x_j}{g}$ the update function $f$ such that $f_j(\vec{x},\vec{r})=g(\vec{x},\vec{r})$ and $f_i(\vec{x},\vec{r})=\vec{x}[i]$ for all $i\neq j$.

We say that an SGS $\mathcal{G}$ is \emph{normalized} if all guards of all transitions in $\mathcal{G}$ are in a disjunctive normal form.

\begin{example}\label{ex:sgs}
Figure~\ref{fig:Q4:SGS} shows an example of stochastic game structure.
Deterministic locations are represented by boxes, angelic locations by triangles, demonic locations by diamonds,
and stochastic locations by circles. Transitions are labelled with update functions, while guards and probabilities
of transitions outgoing from deterministic and stochastic locations, respectively, are given in rounded rectangles
on these transitions. For the sake of succinctness we do not picture tautological guards and identity update functions. Note that the SGS is normalized.
We will describe in Example~\ref{ex:illustrate} how the stochastic game structure shown corresponds to the program described in
Example~\ref{ex:prog}.
\end{example}

\paragraph*{Dynamics of Stochastic Games}

A \emph{configuration} of an SGS $\mathcal{G}$ is a tuple $(\ell,\vec{x})$, where $\ell$ is a location of $\mathcal{G}$ and $\vec{x}$ is a valuation of program variables.
We say that a transition $\tau$ is \emph{enabled} in a configuration $(\ell,\vec{x})$ if $\ell$ is the source location of $\tau$ and in addition, $\vec{x}\models G(\tau)$ provided that $\ell$ is deterministic.

The possible behaviours of the system modelled by $\mathcal{G}$ are represented by \emph{runs} in $\mathcal{G}$. Formally, a \emph{finite path} (or \emph{execution fragment}) in $\mathcal{G}$ is a finite sequence of configurations $(\ell_0,\vec{x}_0)\cdots(\ell_k,\vec{x}_k)$ such that for each $0 \leq i < k$ there is a transition  $(\ell_i,f,\ell_{i+1})$ enabled in $(\ell_i,\vec{x}_i)$ and a valuation $\vec{r}$ of random variables such that $\vec{x}_{i+1} = f(\vec{x}_{i},\vec{r})$. A \emph{run} (or \emph{execution}) of $\mathcal{G}$ is an infinite sequence of configurations whose every finite prefix is a finite path.
A configuration $(\loc,\vec{x})$ is {\em reachable} from the start configuration $(\loc_0,\initval)$
if there is a finite path starting at $(\loc_0,\initval)$ that ends in $(\loc,\vec{x})$.

Due to the presence of non-determinism and probabilistic choices, an SGS $\mathcal{G}$ may exhibit a multitude of possible behaviours. The probabilistic behaviour of $\mathcal{G}$ can be captured by constructing a suitable probability measure over the set of all its runs. However, before this can be done, non-determinism in $\mathcal{G}$ needs to be resolved. To do this, we utilize the standard notion of a \emph{scheduler}.

\begin{definition}
\label{def:schedulers}
An angelic (resp., demonic) scheduler in an SGS $\mathcal{G}$ is a function which assigns to every finite path in $\mathcal{G}$ that ends in an angelic (resp., demonic) configuration $(\ell,\vec{x})$, respectively,
a transition outgoing from~$\ell$.
\end{definition}

Intuitively, we view the behaviour of $\mathcal{G}$ as a game played between two players, angel and demon, with angelic and demonic schedulers representing the strategies of the respective players. That is, schedulers are blueprints for the players that tell them how to play the game.
The behaviour of $\mathcal{G}$ under angelic scheduler $\sigma$ and demonic scheduler $\pi$ can then be intuitively described as follows: The game starts in the initial configuration $(\ell_0,\vec{x}_0)$. In every step $i$, assuming the current configuration to be $(\ell_i,\vec{x}_i)$ the following happens:
%
\begin{compactitem}
\item
A valuation vector $\vec{r}$ for the random variables of $\mathcal{G}$ is sampled according to the distribution $\mathcal{D}$.
\item
A transition $\tau=(\ell_i,f,\ell')$ enabled in $(\ell_i,\vec{x}_i)$ is chosen according to the following rules:
\begin{compactitem}
\item
If $\ell_i$ is angelic (resp., demonic), then $\tau$ is chosen 
deterministically by scheduler
$\sigma$ (resp., $\pi$).
That is, if $\ell_i$ is angelic (resp., demonic) and $c_0c_1\cdots c_i$ is the sequence of configurations observed so far,
then $\tau$ equals 
$\sigma(c_0c_1\cdots c_i)$ (resp., $\pi(c_0c_1\cdots c_i)$).

\item
If $\ell_i$ is probabilistic, then $\tau$ is chosen randomly according to the distribution $\probdist_{\ell_i}$.
\item
If $\ell_i$ is deterministic, then by the definition of an SGS there is exactly one enabled transition outgoing from $\ell_i$, and this transition is chosen as $\tau$.
\end{compactitem}
\item
The transition $\tau$ is traversed and the game enters a new configuration $(\ell_{i+1},\vec{x}_{i+1}) = (\ell',f(\vec{x}_i,\vec{r}))$.
\end{compactitem}

In this way, the players and random choices eventually produce a random run in $\mathcal{G}$. The above intuitive explanation can be formalized by showing that the schedulers $\sigma$ and $\pi$ induce a unique probability measure $\probm^{\sigma,\pi}$ over a suitable $\sigma$-algebra having runs in $\mathcal{G}$ as a sample space. If $\mathcal{G}$ does not have any angelic/demonic locations, there is only one angelic/demonic scheduler (an empty function) that we typically omit from the notation, i.e. if there are no angelic locations we write only $\probm^{\pi}$ etc.

\paragraph*{From Programs to Games}
To every affine probabilistic program $P$ we can assign a stochastic game structure $\mathcal{G}_P$ whose locations correspond to the values of the program counter of $P$ and whose transition relation captures the behaviour of $P$. The game $\mathcal{G}_P$ has the same program and random variables as $P$, with the initial valuation $\vec{x}_0$ of the former and the distribution $\mathcal{D}$ of the latter being specified in the program's preamble. The construction of the state space of $\mathcal{G}_P$ can be described inductively. For each program $P$ the game $\mathcal{G}_P$ contains two distinguished locations, $\ell^{\lin}_{P}$ and $\ell^{\lout}_{P}$, the latter one being always deterministic, that intuitively represent the state of the program counter before and after executing $P$, respectively.
\begin{compactenum}
\item {\em Expression and Skips.}
For $P= {x}{:=}{E}$ where $x$ is a program variable and $E$ is an arithmetic expression, or $P = \textbf{skip}$, the game $\mathcal{G}_P$ consists only 
locations $\ell^{\lin}_P$ and $\ell^{\lout}_P$ (both deterministic) and a transition $(\ell^{\lin}_{P},\assgn{x}{E},\ell^{\lout}_P)$ or $(\ell^{\lin}_{P},\id,\ell^{\lout}_P)$, respectively.

\item {\em Sequential Statements.}
For $P = Q_1;Q_2$ we take the games $\mathcal{G}_{Q_1}$, $\mathcal{G}_{Q_2}$ and join them by identifying the location $\ell^{\lout}_{Q_1}$ with $\ell^{\lin}_{Q_2}$, putting $\ell^{\lin}_{P}=\ell^{\lin}_{Q_1}$ and $\ell^{\lout}_{P}=\ell^{\lout}_{Q_2}$.

\item {\em While Statements.}
For $P = \textbf{while $\phi$ do }Q \textbf{ od}$ we add a new deterministic location $\ell^{\lin}_{P}$ which we identify with $\ell^{\lout}_{Q}$, a new deterministic location $\ell^{\lout}_{P}$, and transitions $\tau=(\ell^{\lin}_{P},\id,\ell^{\lin}_{Q})$, $\tau'=(\ell^{\lin}_{P},\id,\ell^{\lout}_{P})$ such that $G(\tau)=\phi$ and $G(\tau')=\neg\phi$.

\item {\em If Statements.}
Finally, for $P = \textbf{if $\mathit{ndb}$ then }Q_1 \textbf{ else } Q_2 \textbf{ fi}$ we add a new location $\ell^{\lin}_{P}$ together with two transitions $\tau_1 = (\ell^{\lin}_{P},\id,\ell^{\lin}_{Q_1})$, $\tau_2 = (\ell^{\lin}_{P},\id,\ell^{\lin}_{Q_2})$, and we identify the locations  $\ell^{\lout}_{Q_1}$ and $\ell^{\lout}_{Q_1}$ with $\ell^{\lout}_{P}$. In this case the newly added location $\ell^\lin_{P}$ is angelic/demonic if and only if $ndb$ is the keyword '\textbf{angel}'/'\textbf{demon}', respectively. If $\mathit{ndb}$ is of the form $\textbf{prob($p$)}$, the location $\ell^\lin_{P}$ is probabilistic with $\probdist_{\ell^\lin_{P}}(\tau_1)=p$ and $\probdist_{\ell^\lin_{P}}(\tau_2)=1-p$. Otherwise (i.e. if $\mathit{ndb}$ is a propositionally linear predicate), $\ell^\lin_{P}$ is a deterministic location with $G(\tau_1)=\mathit{ndb}$ and $G(\tau_2)=\neg \mathit{ndb}$.
\end{compactenum}
Once the game $\mathcal{G}_P$ is constructed using the above rules, we put $G(\tau)=\textit{true}$ for all transitions $\tau$ outgoing from deterministic locations whose guard was not set in the process, and finally we add a self-loop on the location $\ell^{\lout}_P$. This ensures that the assumptions in Definition~\ref{def:stochgame} are satisfied.
Furthermore note that for SGS obtained for a program $P$, since the only
branching are conditional branching, every location $\loc$ has at most two
successors $\loc_1,\loc_2$.

\begin{example}\label{ex:illustrate}
We now illustrate step by step how the SGS of Example~\ref{ex:sgs} corresponds to
the program of Example~\ref{ex:prog}.
We first consider the statements $Q_1$ and $Q_2$ (Figure~\ref{fig:Q1Q2}), and show the
corresponding SGSs in Figure~\ref{fig:Q1Q2:SGS}.
Then consider the statement block $Q_3$ which is a probabilistic choice between
$Q_1$ and $Q_2$  (Figure~\ref{fig:Q3}).
The corresponding SGS  (Figure~\ref{fig:Q3:SGS}) is obtained from the previous two
SGSs as follows: we consider a probabilistic start location where there is a probabilistic
branch to the start locations of the SGSs of $Q_1$ and $Q_2$, and the SGS ends
in a location with only self-loop.
Finally, we consider the whole program as $Q_4$ (Figure~\ref{fig:Q4}),
and the corresponding SGS in  (Figure~\ref{fig:Q4:SGS}).
The SGS is obtained from SGS for $Q_3$, with the self-loop replaced by a transition
back to the probabilistic location (with guard $x\geq 0$), and an edge to the final location
(with guard $x<0$).
The start location of the whole program is a new location, with transition labeled
$x:=0$, to the start of the while loop location.
We label the locations in Figure~\ref{fig:Q4:SGS} to refer to them later.
\lstset{language=affprob}
\lstset{tabsize=3}
\newsavebox{\figa}
\begin{lrbox}{\figa}
\begin{lstlisting}[mathescape]
$Q_1$:	if angel then $x:=x+1$ else $x:=x-1$ fi	
$Q_2$:	if demon then $x:=x+1$ else $x:=x-1$ fi
\end{lstlisting}
\end{lrbox}
\begin{figure}[h]
\centering
\usebox{\figa}
\caption{Programs $Q_1$ and $Q_2$}
\label{fig:Q1Q2}
\end{figure}

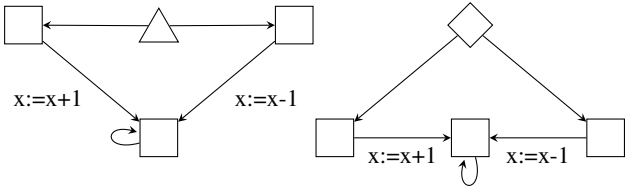
\begin{figure}[h]
\centering
\begin{tikzpicture}[x=1.8cm]
\node[det] (fin1) at (0,-1.5)  {};
\node[ang] (angel) at (0,0) {};
\node[det] (fin2) at (2.3,-1.5)  {};
\node[dem] (demon) at (2.3,0)  {};
\node[det] (aif) at (-1,0) {};
\node[det] (aelse) at (1,0) {};
\node[det] (dif) at (1.3,-1.5) {};
\node[det] (delse) at (3.3,-1.5) {};

\draw[tran] (angel.155) -- (aif);
\draw[tran] (angel.25) -- (aelse);
\draw[tran] (demon) -- (dif);
\draw[tran] (demon) -- (delse);

\path (fin1) edge [tran, loop left] (fin1);
\path (fin2) edge [tran, loop below] (fin2);
\draw[tran] (dif) -- node[auto,swap] {x:=x+1} (fin2);
\draw[tran] (delse) -- node[auto] {x:=x-1} (fin2);
\draw[tran] (aif) -- node[auto, swap] {x:=x+1} (fin1);
\draw[tran] (aelse) to node[auto] {x:=x-1} (fin1);
\end{tikzpicture}
\caption{SGSs for $Q_1$ (left) and $Q_2$ (right)}
\label{fig:Q1Q2:SGS}
\end{figure}

\newsavebox{\figb}
\begin{lrbox}{\figb}
\begin{lstlisting}[mathescape]
$Q_3$:	if prob(0.6) then $Q_1$ else $Q_2$ fi
\end{lstlisting}
\end{lrbox}
\begin{figure}[h]
\centering
\usebox{\figb}
\caption{Program $Q_3$}
\label{fig:Q3}
\end{figure}
\begin{figure}[h]
\centering
\begin{tikzpicture}[x = 1.8cm]
\node[det] (while) at (1.5,-1.5)  {};
\node[ran] (prob) at (3.7,-1.5) {};
\node[ang] (angel) at (3,0) {};
\node[dem] at (3,-3) (demon) {};
\node[det] (aif) at (2,-0.075) {};
\node[det] (aelse) at (3,-0.8) {};
\node[det] (dif) at (2,-3) {};
\node[det] (delse) at (3,-2.2) {};

\draw[tran] (angel) -- (aif);
\draw[tran] (angel) -- (aelse);
\draw[tran] (demon) -- (dif);
\draw[tran] (demon) -- (delse);

\path (while) edge [tran, loop left] (while);
\draw[tran] (prob) -- node[font=\scriptsize,draw, fill=white, rectangle,pos=0.3, inner sep = 1pt] {$\frac{6}{10}$} (prob|-angel) -- (angel);
\draw[tran] (prob) -- node[font=\scriptsize,draw, fill=white, rectangle,pos=0.3, inner sep = 1pt] {$\frac{4}{10}$} (prob|-demon) --(demon);
\draw[tran] (dif) --  (demon-|while) -- node[auto, pos=0.2] {x:=x+1} (while);
\draw[tran] (aif) --  (angel-|while) -- node[auto, swap, pos=0.2] {x:=x+1} (while);
\draw[tran] (delse) -- node[auto] {x:=x-1} (delse-|while.305) --  (while.-55);
\draw[tran] (aelse) -- node[auto, swap] {x:=x-1} (aelse-|while.55) --  (while.55);
\end{tikzpicture}
\caption{SGS of $Q_3$}
\label{fig:Q3:SGS}
\end{figure}
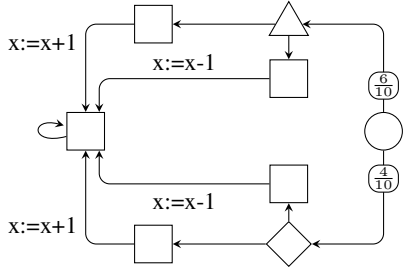
\newsavebox{\figc}
\begin{lrbox}{\figc}
\begin{lstlisting}[mathescape]
$Q_4$:	$x:=0$; while ($x\ge 0$) do $Q_3$ od
\end{lstlisting}
\end{lrbox}
\begin{figure}[h]
\centering
\usebox{\figc}
\caption{Program $Q_4$}
\label{fig:Q4}
\end{figure}
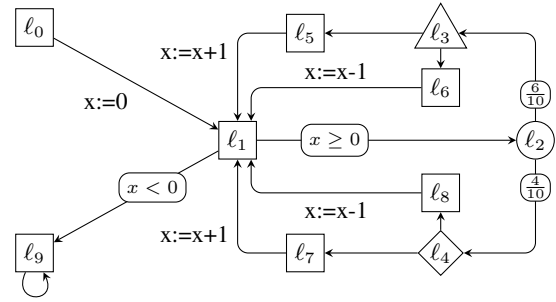
\begin{figure}[h]
\centering
\begin{tikzpicture}[x = 1.8cm]
%

\node[det] (while) at (1.5,-1.5)  {$\loc_1$};
\node[ran] (prob) at (3.7,-1.5) {$\loc_2$};
\node[ang] (angel) at (3,0) {$\loc_3$};
\node[dem] at (3,-3) (demon) {$\loc_4$};
\node[det] (aif) at (2,-0.075) {$\loc_5$};
\node[det] (aelse) at (3,-0.8) {$\loc_6$};
\node[det] (dif) at (2,-3) {$\loc_7$};
\node[det] (delse) at (3,-2.2) {$\loc_8$};
\node[det] (det1) at (0,0) {$\loc_0$};
\node[det] (fin) at (0,-3) {$\loc_9$};
%
\draw[tran] (det1) to node[auto, swap] {x:=0} (while);
\draw[tran] (while) to node[font=\scriptsize,draw, fill=white, rectangle,pos=0.4] {$x<0$} (fin);
\draw[tran, loop, looseness = 5, in =-65, out = -115] (fin) to (fin);
\draw[tran] (while) to node[font=\scriptsize,draw, fill=white, rectangle,pos=0.3] {$x\geq 0$} (prob);

\draw[tran] (angel) -- (aif);
\draw[tran] (angel) -- (aelse);
\draw[tran] (demon) -- (dif);
\draw[tran] (demon) -- (delse);

\draw[tran] (prob) -- node[font=\scriptsize,draw, fill=white, rectangle,pos=0.3, inner sep = 1pt] {$\frac{6}{10}$} (prob|-angel) -- (angel);
\draw[tran] (prob) -- node[font=\scriptsize,draw, fill=white, rectangle,pos=0.3, inner sep = 1pt] {$\frac{4}{10}$} (prob|-demon) --(demon);
\draw[tran] (dif) --  (demon-|while) -- node[auto, pos=0.2] {x:=x+1} (while);
\draw[tran] (aif) --  (angel-|while) -- node[auto, swap, pos=0.2] {x:=x+1} (while);
\draw[tran] (delse) -- node[auto] {x:=x-1} (delse-|while.305) --  (while.-55);
\draw[tran] (aelse) -- node[auto, swap] {x:=x-1} (aelse-|while.55) --  (while.55);
\end{tikzpicture}
\caption{SGS of $Q_4$}
\label{fig:Q4:SGS}
\end{figure}
\end{example}

\subsection{Qualitative and Quantitative Termination Questions}
We consider the most basic notion of liveness, namely {\em termination},
for probabilistic programs, and present the  relevant {\em qualitative} and
{\em quantitative} questions.

\paragraph*{Qualitative question.}
We consider the two basic qualitative questions, namely, almost-sure termination (i.e., termination with probability~$1$) and finite expected termination time.
We formally define them below.

Given a program $P$, let $\mathcal{G}_P$ be the associated SGS.
A run $\rho$ is \emph{terminating} if it reaches a configuration in which the
location is $\ell^{\lout}_{P}$.
Consider the random variable $T$ which to every run $\rho$ in $\mathcal{G}_P$
assigns the first point in time in which a configuration with the location $\ell^{\lout}_{P}$ is encountered,
and if the run never reaches such a configuration, then the value assigned is $\infty$.

\begin{definition}[Qualitative termination questions]
Given a program $P$ and its associated normalized SGS $\mathcal{G}_P$, we consider the following
two questions:
\begin{compactenum}
\item {\em Almost-Sure Termination.} The program is {\em almost-surely (a.s.)} terminating if
there exists an angelic scheduler (called {\em a.s terminating}) $\sigma$ such that for all demonic
schedulers $\pi$ we have
$\probm^{\sigma,\pi}\left( \{\rho\mid\rho \text{ is terminating}\,\}\right) = 1;$ or equivalently,
$\probm^{\sigma,\pi}(T<\infty)=1$.

\item {\em Finite Termination.}
The program $P$ is \emph{finitely terminating} (aka positively almost-sure terminating) if there
exists an angelic scheduler $\sigma$ (called {\em finitely terminating})
such that for all demonic schedulers $\pi$ it holds that $\expv^{\sigma,\pi}[T]<\infty.$
\end{compactenum}

\end{definition}

Note that for all angelic schedulers $\sigma$ and demonic schedulers $\pi$ we
have $\expv^{\sigma,\pi}[T]<\infty$ implies $\probm^{\sigma,\pi}(T<\infty)=1$,
however, the converse does not hold in general.
In other words, finitely terminating implies a.s. terminating, but a.s.
termination does not imply finitely termination.

\begin{definition}[Quantitative termination questions]
\label{def:quanttq}
Given a program $P$ and its associated normalized SGS $\mathcal{G}_P$, we consider the following
notions:
\begin{compactenum}
\item {\em Expected Termination Time.} The {\em expected termination} time of $P$ is
$\Eval(P) = \inf_{\sigma}\sup_{\pi}\expv^{\sigma,\pi}[T]. $

\item {\em Concentration bound.} A bound $B$ is a {\em concentration bound} if there exists two positive constants $c_1$ and $c_2$ such that for all $x\geq B$, we have
$\Thr(P,x) \leq c_1 \cdot \exp(-c_2\cdot x)$, where $\Thr(P,x)=\inf_{\sigma} \sup_{\pi} \probm^{\sigma,\pi}(T > x)$ (i.e., the probability that the termination time exceeds $x \geq B$ decreases exponentially in $x$).
\end{compactenum}
\end{definition}

Note that we assume that an SGS $\mathcal{G}_P$ for a program $P$ is already given in a normalized form, as our algorithms assume that all guards in $\mathcal{G}_P$ are in DNF.
In general, converting an SGS into a normalized SGS incurs an exponential blow-up as this is the worst-case blowup when converting a formula into DNF. However, we note that for programs $\mathcal{P}$ that contain only \emph{simple guards}, i.e. guards that are either conjunctions or disjunctions of linear constraints, a normalized game $\mathcal{G}_P$ can be easily constructed in polynomial time using de Morgan laws. In particular, we stress that all our hardness results hold already for programs with simple guards, so they do not rely on the requirement that $\mathcal{G}_P$ must be normalized.


\section{The Class \LRAPP}\label{sec:lrapp}
For probabilistic programs a very powerful technique to establish termination
is based on ranking supermartingales.
The simplest form of ranking supermartingales are the {\em linear} ranking ones.
In this section we will consider the class of \APP 's for which linear
ranking supermartingales exist, and refer to it as \LRAPP.
Linear ranking supermartingales have been considered for probabilistic
programs without any types of non-determinism~\cite{SriramCAV}.
We show how to extend the approach in the presence of two types of
non-determinism.
We also show that in \LRAPP~ we have that a.s. termination coincides with
finite termination
(i.e., in contrast to the general case where a.s. termination might not imply
finite-termination, for the well-behaved class of \LRAPP~ we have
a.s. termination implies finite termination).
We first present the general notion of ranking supermartingales, and
will establish their role in qualitative termination.

\begin{definition}[Ranking Supermartingales~\cite{HolgerPOPL}]\label{def:rsm}
A discrete-time stochastic process $\{X_n\}_{n\in\mathbb{N}}$ wrt a
filtration $\{\mathcal{F}_n\}_{n\in\mathbb{N}}$ is a \emph{ranking supermartingale} (RSM)
if there exists $K<0$ and $\epsilon>0$ such that  for all $n\in\mathbb{N}$, $\expv(|X_n|)$ exists and it holds almost surely
(with probability~1) that
\[
X_n\ge K\mbox{ and }\expv(X_{n+1}\mid \mathcal{F}_n)\le X_n-\epsilon\cdot\mathbf{1}_{X_n\ge 0}\enskip,
\]
where $\expv(X_{n+1}\mid \mathcal{F}_n)$ is the conditional expectation of $X_{n+1}$ given the $\sigma$-algebra $\mathcal{F}_n$
(cf.~\cite[Chapter~9]{probabilitycambridge}).
\end{definition}

In following proposition we establish (with detailed proof in the appendix)
the relationship between RSMs and certain notion of termination time.

\begin{proposition}\label{prop:rsm}
Let $\{X_n\}_{n\in\mathbb{N}}$ be an RSM wrt filtration $\{\mathcal{F}_n\}_{n\in\mathbb{N}}$ and let numbers $K,\epsilon$ be as in Definition~\ref{def:rsm}.
Let $Z$ be the random variable defined as
$Z:=\min\{n\in\mathbb{N}\mid X_n<0\}$;
which denotes the first time $n$ that the RSM $X_n$ drops below~0.
Then $\probm(Z<\infty)=1$ and $\expv(Z)\le\frac{\expv(X_1)-K}{\epsilon}$.
\end{proposition}

\begin{remark}\label{rmk:scaling}
WLOG we can consider that the constants $K$ and $\epsilon$ in Definition~\ref{def:rsm} satisfy that
$K\le -1$ and $\epsilon\ge 1$, as an RSM can be scaled by a positive scalar to ensure that
$\epsilon$ and the absolute value of $K$ are sufficiently large.
\end{remark}

For the rest of the section we fix an affine probabilistic program $P$
and let $\mathcal{G}_P=(\locs,(\pvars,\rvars),\ell_0,\vec{x}_0,\transitions,\probdist,\guards)$
be its associated SGS.
We fix the filtration $\{\mathcal{F}_n\}_{n\in\mathbb{N}}$ such that each $\mathcal{F}_n$ is the smallest $\sigma$-algebra on runs that
makes all random variables in $\{\theta_j\}_{1\le j\le n},\{\overline{x}_{k,j}\}_{1\le k\le |\pvars|, 1\le j\le n}$ measurable,
where $\theta_j$ is the random variable representing the location at the $j$-th step (note that each location can be deemed as a natural number.),
and $\overline{x}_{k,j}$ is the random variable representing the value of the program variable $x_{k}$ at the $j$-th step.

To introduce the notion of linear ranking supermartingales,
we need the notion of \emph{linear invariants} defined as follows.

\begin{definition}[Linear Invariants]
A \emph{linear invariant} on $\pvars$ is a function $\inv$ assigning a finite set of non-empty
linear assertions on $\pvars$ to each location of $\mathcal{G}_P$ such that for all configurations $(\loc,\vec{x})$ reachable from $(\loc_0,\initval)$ in $\mathcal{G}_P$ it holds that $\vec{x}\in \bigcup I(\loc)$.
\end{definition}

Generation of linear invariants can be done through abstract interpretation~\cite{DBLP:conf/popl/CousotC77}, as adopted
in~\cite{SriramCAV}.
We first extend the notion of pre-expectation~\cite{SriramCAV} to both angelic and demonic non-determinism.

\begin{definition}[Pre-Expectation]
Let $\eta:L\times\mathbb{R}^{|\pvars|}\rightarrow\mathbb{R}$ be a function.
The function $\mathrm{pre}_\eta:L\times\mathbb{R}^{|\pvars|}\rightarrow\mathbb{R}$ is defined by:
\begin{itemize}\itemsep1pt \parskip0pt \parsep0pt
\item $\mathrm{pre}_\eta(\loc,\mathbf{x}):=\sum_{(\loc,id,\loc')\in\transitions} Pr_{\loc}\left(\loc,id,\loc'\right)\cdot \eta\left(\loc',\mathbf{x}\right)$ if $\loc$ is a probabilistic location;
\item $\mathrm{pre}_\eta(\loc,\mathbf{x}):=\max_{(\loc,id,\ell')\in\transitions}\eta(\loc',\mathbf{x})$ if $\loc$ is a demonic location;
\item $\mathrm{pre}_\eta(\loc,\mathbf{x}):=\min_{(\loc,id,\ell')\in\transitions}\eta\left(\loc',\mathbf{x}\right)$ if $\loc$ is an angelic location;
\item $\mathrm{pre}_\eta(\loc,\mathbf{x}):=\eta(\loc',\expv_{\rvars}\left(f(\vec{x},\mathbf{r})\right))$ if $\loc$ is a deterministic location, $(\loc,f,\loc')\in\transitions$ and $\mathbf{x}\in G(\loc,f,\loc')$, where $\expv_{\rvars}\left(f(\vec{x},\mathbf{r})\right)$ is the expected value of $f(\vec{x},\cdot)$. 
\end{itemize}
\end{definition}
Intuitively, $\mathrm{pre}_\eta(\loc,\mathbf{x})$ is the one-step optimal expected value of $\eta$ from the configuration $(\loc,\mathbf{x})$.
In view of Remark~\ref{rmk:scaling}, the notion of linear ranking supermartingales is now defined as follows.

\begin{definition}[Linear Ranking-Supermartingale Maps]~\label{def:lrsm}
A \emph{linear ranking-supermartingale map} (LRSM) wrt a linear invariant $I$ for $\mathcal{G}_P$ is a function $\eta: \locs\times\mathbb{R}^{|\pvars|}\rightarrow\mathbb{R}$ such that the following conditions (C1-C4) hold: there exist $\epsilon \geq 1$ and $K,K'\leq -1$ such that for all $\loc\in\locs$ and all $\vec{x}\in\mathbb{R}^{|\pvars|}$, we have
\begin{itemize}\itemsep1pt \parskip0pt \parsep0pt
\item {\em C1:} the function $\eta(\loc,\centerdot):\mathbb{R}^{|\pvars|}\rightarrow\mathbb{R}$
is linear over the program variables $\pvars$;
\item {\em C2:} if $\loc\ne \loc_P^\lout$ and $\vec{x}\in \bigcup I(\loc)$, then $\eta(\loc,\mathbf{x})\ge 0$;
\item {\em C3:} if $\loc=\loc_P^\lout$ and $\vec{x}\in \bigcup I(\loc)$, then $K'\le \eta(\loc, \vec{x})\le K$;
\item {\em C4:}
\[
\left(\loc\ne\loc_P^\lout\wedge\vec{x}\in \bigcup I(\loc)\right)\rightarrow\mathrm{pre}_\eta(\loc,\mathbf{x})\le\eta(\loc,\mathbf{x})-\epsilon \enskip.
\]
\end{itemize}
\end{definition}
We refer to the above conditions as follows: C1 is the {\em linearity} condition;
C2 is the {\em non-terminating non-negativity} condition, which specifies that for every non-terminating
location the RSM is non-negative;
C3 is the {\em terminating negativity} condition, which specifies that in the terminating
location the RSM is negative (less than $-1$) and lowerly bounded;
C4 is the {\em supermartingale difference} condition which is intuitively
related to the $\epsilon$ difference in the RSM definition
(cf Definition~\ref{def:rsm}).

\begin{remark}~\label{rmk:linearconstraintsptime} 
In~\cite{SriramCAV}, the condition C3 is written as $K'\le \eta(\loc, \vec{x})<0$ and is handled by Motzkin's Transposition Theorem, resulting in possibly quadratic constraints.
Here, we replace $\eta(\loc, \vec{x})<0$ equivalently with $\eta(\loc, \vec{x})\le K$ which allows one to obtain linear constraints through Farkas' linear assertion,
where the equivalence follows from the fact that maximal value of a linear program can be attained if it is finite.
This is crucial to our PTIME result over programs with at most demonic non-determinism.
\end{remark}

Informally, LRSMs extend \emph{linear expression maps} defined in~\cite{SriramCAV}
with both angelic and demonic non-determinism.
The following theorem establishes the soundness of LRSMs.

\begin{theorem}
\label{thm:supermartingale-correctness}
If there exists an LRSM $\eta$ wrt $I$ for $\mathcal{G}_P$, then
\begin{compactenum}
\item $P$ is a.s. terminating; and
\item $\Eval(P)\leq \frac{\eta(\initloc,\initval)-K'}{\epsilon}$. In particular, $\Eval(P)$ is finite.
\end{compactenum}
\end{theorem}
\smallskip\noindent{\em Key proof idea.}
Let $\eta$ be an LRSM, wrt a linear invariant $I$ for $\mathcal{G}_P$.
Let $\sigma$ be the angelic scheduler whose decisions optimize the value of $\eta$
at the last configuration of any finite path, represented by
\[
\sigma(\loc,\vec{x})=\mathrm{argmin}_{(\loc,f,\loc')\in\transitions}\eta(\loc',\vec{x})
\]
for all end configurations $(\loc,\vec{x})$ such that
$\loc\in L_A$ and $\vec{x}\in\mathbb{R}^{|\pvars|}$.
Fix any demonic strategy $\pi$.
Let the stochastic process $\{X_n\}_{n\in\mathbb{N}}$ be defined by: $X_n(\omega):=\eta(\theta_n(\omega),\{\overline{x}_{k,n}(\omega)\}_{k})$.
We show that $\{X_n\}_{n\in\mathbb{N}}$ is an RSM, and then use  Proposition~\ref{prop:rsm}
to obtain the desired result (detailed proof in appendix).

\begin{remark}\label{rmk:scheduler}
Note that the proof of Theorem~\ref{thm:supermartingale-correctness} also provides a way to synthesize
an angelic scheduler, given the LRSM, to ensure that the expected termination time is finite
(as our proof gives an explicit construction of such a scheduler).
Also note that the result provides an upper bound, which we denote as $\UB(P)$, on
$\Eval(P)$.
\end{remark}

\smallskip\noindent{\em The class \LRAPP.}
The class \LRAPP{} consists of all \APP's for which
there exists a linear invariant $I$ such that an LRSM exists w.r.t $I$ for $\mathcal{G}_P$.
It follows from Theorem~\ref{thm:supermartingale-correctness} that programs in \LRAPP{} terminate almost-surely, and have finite expected termination time.

\section{\LRAPP: Qualitative Analysis}
In this section we study the computational problems related to \LRAPP.
We consider the following basic
computational questions regarding {\em realizability} and {\em synthesis.}

\smallskip\noindent{\bf \LRAPP\ realizability and synthesis.}
Given an \APP $P$ with its normalized SGS $\mathcal{G}_P$ and a linear invariant $I$,
we consider the following questions:
\begin{compactenum}
\item {\em \LRAPP\ realiazability.} Does there exist an LRSM wrt $I$ for $\mathcal{G}_P$?

\item {\em \LRAPP\ synthesis.} If the answer to the realizability question is yes, then
construct a witness LRSM.
\end{compactenum}
Note that the existence of an LRSM implies almost-sure and finite-termination
(Theorem~\ref{thm:supermartingale-correctness}), and presents
affirmative  answers to
the qualitative questions.
We establish the following result.

\begin{theorem}\label{thm:mainalgorithm}
The following assertions hold:
\begin{compactenum}
\item The \LRAPP\ realizability and synthesis problems for programs in \APP s
can be solved in $\PSPACE$, by solving a set of quadratic constraints.

\item For programs in \APP s with only demonic non-determinism, the \LRAPP\ realizability
and synthesis problems can be solved in polynomial time, by solving
a set of linear constraints.

\item Even for programs in \APP s with simple guards, only angelic non-determinism, and no probabilistic
choice, the \LRAPP\ realizability problem is $\NP$-hard.
\end{compactenum}

\end{theorem}

\smallskip\noindent{\em Discussion and organization.}
The significance of our result is as follows: it presents a practical approach
(based on quadratic constraints for general \APP s, and linear constraints for
\APP s with only demonic non-determinism) for the problem, and on the other
hand it shows a sharp contrast in the complexity between the case with angelic
non-determinism vs demonic non-determinism ($\NP$-hard vs PTIME).
In Section~\ref{subsec:algo} we present an algorithm to establish the first
two items, and then establish the hardness result in
Section~\ref{subsec:hard1}.

\subsection{Algorithm and Upper Bounds}\label{subsec:algo}

\noindent{\em Solution overview.}
Our algorithm is based on an encoding of the conditions (C1--C4) for an LRSM into a set of
universally quantified formulae.
Then the universally quantified formulae are translated to existentially quantified
formulae, and the key technical machineries are Farkas' Lemma and  Motzkin's
Transposition Theorem (which we present below).

\begin{theorem}[Farkas' Lemma~\cite{FarkasLemma,SchrijverPolyhedra}]
Let $\mathbf{A}\in\mathbb{R}^{m\times n}$, $\mathbf{b}\in\mathbb{R}^m$, $\mathbf{c}\in\mathbb{R}^{n}$ and $d\in\mathbb{R}$.
Assume that $\{\mathbf{x}\mid \mathbf{A}\mathbf{x}\le \mathbf{b}\}\ne\emptyset$.
Then
\[
\{\mathbf{x}\mid \mathbf{A}\mathbf{x}\le \mathbf{b}\}\subseteq \{\mathbf{x}\mid \mathbf{c}^{\mathrm{T}}\mathbf{x}\le d\}
\]
iff there exists $\mathbf{y}\in\mathbb{R}^m$ such that $\mathbf{y}\ge \mathbf{0}$, $\mathbf{A}^\mathrm{T}\mathbf{y}=\mathbf{c}$ and $\mathbf{b}^{\mathrm{T}}\mathbf{y}\le d$.
\end{theorem}

\smallskip\noindent{\bf Farkas' linear assertion $\Phi$.}
Farkas' Lemma transforms the inclusion testing of non-strict systems of linear inequalities into the emptiness problem.
For the sake of convenience, given a polyhedron $H=\{\mathbf{x}\mid \mathbf{A}x\le \mathbf{b}\}$ with $\mathbf{A}\in\mathbb{R}^{m\times n},\mathbf{b}\in\mathbb{R}^m$ and $\mathbf{c}\in\mathbb{R}^{n},d\in\mathbb{R}$,
we define the linear assertion $\Phi[H,\mathbf{c},d](\xi)$ (which we refer to as Farkas' linear assertion)
for Farkas' Lemma by
\[
\Phi[H,\mathbf{c},d](\xi):=\xi\ge \mathbf{0}\wedge \mathbf{A}^\mathrm{T}\xi=\mathbf{c}\wedge\mathbf{b}^{\mathrm{T}}\xi\le d\enskip,
\]
where $\xi$ is a column-vector variable of dimension $n$.
Moreover, let 
\[
\Phi[\mathbb{R}^n,\mathbf{c},d](\xi):=\vec{c}=\vec{0}\wedge d\ge 0\enskip.
\]
Note that $\mathbb{R}^n\subseteq \{\vec{x}\mid \vec{c}^{\mathrm{T}}\vec{x}\le d\}$ iff
$\vec{c}=\vec{0}$ and $d\ge 0$.

Below we show (proof in the appendix) that Farkas' Lemma can be slightly
extended to strict inequalities.

\begin{lemma}\label{lemm:closedness}
Let $\mathbf{A}\in\mathbb{R}^{m\times n}$, $\mathbf{B}\in\mathbb{R}^{k\times n}$, $\mathbf{b}\in\mathbb{R}^{m}$, $\mathbf{d}\in\mathbb{R}^k$, $\mathbf{c}\in\mathbb{R}^{n}$ and $d\in\mathbb{R}$.
Let
$\mathbf{Z}_{<} = \{\mathbf{x}\mid \mathbf{A}\mathbf{x}\le \mathbf{b}\wedge \mathbf{B}\mathbf{x}< \mathbf{d}\}$ and
$\mathbf{Z}_{\leq} =\{\mathbf{x}\mid \mathbf{A}\mathbf{x}\le \mathbf{b}\wedge \mathbf{B}\mathbf{x}\leq \mathbf{d}\}$.
Assume that $\mathbf{Z}_{<} \neq \emptyset$.
Then for all closed subsets $H\subseteq\mathbb{R}^{|\pvars|}$ we have that
$\mathbf{Z}_{<} \subseteq H$ implies $\mathbf{Z}_{\leq} \subseteq H$.
\end{lemma}

\begin{remark}
Lemma~\ref{lemm:closedness} is crucial to ensure that our approach can be done in polynomial time when $P$ does not involve angelic non-determinism.
\end{remark}

The following theorem, entitled Motzkin's Transposition Theorem, handles general systems of linear inequalities with strict inequalities.

\begin{theorem}[Motzkin's Transposition Theorem~\cite{Motzkin}]
Let $\mathbf{A}\in\mathbb{R}^{m\times n}$, $\mathbf{B}\in\mathbb{R}^{k\times n}$ and
$\mathbf{b}\in\mathbb{R}^m,\mathbf{c}\in\mathbb{R}^k$.
Assume that $\{\mathbf{x}\in\mathbb{R}^n\mid \mathbf{A}\mathbf{x}\le \mathbf{b}\}\ne\emptyset$.
Then
\[
\{\mathbf{x}\in\mathbb{R}^n\mid \mathbf{A}\mathbf{x}\le \mathbf{b}\}\cap\{\mathbf{x}\in\mathbb{R}^n\mid \mathbf{B}\mathbf{x}<\mathbf{c}\}=\emptyset
\]
iff there exist $\mathbf{y}\in\mathbb{R}^m$ and $\mathbf{z}\in\mathbb{R}^k$ such that $\mathbf{y},\mathbf{z}\ge\mathbf{0}$, $\mathbf{1}^\mathrm{T}\cdot\mathbf{z}>0$,
$\mathbf{A}^{\mathrm{T}}\mathbf{y}+\mathbf{B}^{\mathrm{T}}\mathbf{z}=\mathbf{0}$ and $\mathbf{b}^{\mathrm{T}}\mathbf{y}+\mathbf{c}^{\mathrm{T}}\mathbf{z}\le 0$.
\end{theorem}

\begin{remark}
The version of Motzkin's Transposition Theorem here is a simplified one obtained by taking into account the assumption $\{\mathbf{x}\in\mathbb{R}^n\mid \mathbf{A}\mathbf{x}\le \mathbf{b}\}\ne\emptyset$.
\end{remark}

\smallskip\noindent{\bf Motzkin assertion $\Psi$.}
Given a polyhedron $H=\{\mathbf{x}\mid \mathbf{A}x\le \mathbf{b}\}$ with
$\mathbf{A}\in\mathbb{R}^{m\times n},\mathbf{b}\in\mathbb{R}^m$ and $\mathbf{B}\in\mathbb{R}^{k\times n},\mathbf{c}\in\mathbb{R}^k$, we define
assertion $\Psi[H,\mathbf{B},\mathbf{c}](\xi,\zeta)$ (which we refer as Motzkin assertion)
for Motzkin's Theorem by
\begin{align*}
&\Psi[H,\mathbf{B},\mathbf{c}](\xi,\zeta):=\xi\ge\mathbf{0}\wedge\zeta\ge\mathbf{0}\wedge\\
&\quad\mathbf{1}^\mathrm{T}\cdot\zeta>0\wedge\mathbf{A}^{\mathrm{T}}\xi+\mathbf{B}^{\mathrm{T}}\zeta=\mathbf{0}\wedge\mathbf{b}^{\mathrm{T}}\xi+\mathbf{c}^{\mathrm{T}}\zeta\le 0\enskip,
\end{align*}
where $\xi$ (resp. $\zeta$) is an $m$-dimensional (resp. $k$-dimensional) column-vector variable.
Note that if all the parameters $H,\mathbf{B}$ and $\mathbf{c}$ are constant, then the assertion is linear,
however, in general the assertion is quadratic.

\smallskip\noindent{\em Handling emptiness check.}
The results described till now on linear inequalities require that certain sets defined
by linear inequalities are nonempty.
The following lemma presents a way to detect whether such a set is empty (proof in the appendix).

\begin{lemma}\label{lemm:linearemptiness}
Let $\mathbf{A}\in\mathbb{R}^{m\times n}$, $\mathbf{B}\in\mathbb{R}^{k\times n}$, $\mathbf{b}\in\mathbb{R}^{m}$ and $\mathbf{d}\in\mathbb{R}^k$.
Then all of the following three problems can be decided in polynomial time in the binary encoding of $\mathbf{A},\mathbf{B},\mathbf{b},\mathbf{d}$:
\begin{enumerate}\itemsep1pt \parskip0pt \parsep0pt
\item $\{\mathbf{x}\mid \mathbf{A}\mathbf{x}\le \mathbf{b}\}\stackrel{?}{=}\emptyset$;
\item $\{\mathbf{x}\mid \mathbf{A}\mathbf{x}\le \mathbf{b}\wedge \mathbf{B}\mathbf{x}< \mathbf{d}\}\stackrel{?}{=}\emptyset$;
\item $\{\mathbf{x}\mid \mathbf{B}\mathbf{x}< \mathbf{d}\}\stackrel{?}{=}\emptyset$.
\end{enumerate}
\end{lemma}

Below we fix an input \APP~$P$.

\begin{notations}[Notations for Our Algorithm]\label{notation}
Our algorithm for LRSM realizability and synthesis, which we call \LRSMSynth, is notationally heavy.
To present the algorithm succinctly we will use the following notations (that will be repeatedly used in the algorithm).
\begin{compactenum}
\item {\em $H_{k,\loc}$:} We let $I(\loc)=\bigcup_{k=1}^{k_{\loc}} \{H_{k,\loc}\}$, where each $H_{k,\loc}$ is a satisfiable linear assertion.

\item {\em $G(\tau)$: }For each transition $\tau$ (in the SGS $\mathcal{G}_P$), we deem the propositionally linear predicate $G(\tau)$ also as a set of linear assertions whose members are exactly the conjunctive sub-clauses of $G(\tau)$.

\item {\em $\NS{H}$:} For each linear assertion $H$, we define $\NS{H}$ as follows:
(i)~if $H=\emptyset$, then $\NS{H}:=\emptyset$; and
(ii)~otherwise, the polyhedron $\NS{H}$ is obtained by changing each appearance of `$<$'
(resp. `$>$') by `$\le$' (resp. `$\ge$') in $H$. 
In other words, $\NS{H}$ is the non-strict inequality version of $H$.

\item {\em $\transitions_{\loc}$:} We define
\[
\transitions_\loc:=\{(\loc',f,\loc'')\in\transitions\mid \loc'=\loc\}
\]
to be the set of transitions from $\loc\in L$.

\item {\em $\Op(\loc)$:} For a location $\loc$ we call $\Op(\loc)$ the following open sentence:
\[
\forall\vec{x}\in \bigcup I(\loc).\left[\mathrm{pre}_\eta(\loc,\mathbf{x})\le\eta(\loc,\mathbf{x})-\epsilon\right];
\]
which specifies the condition C4 for LRSM.

\item We will consider $\left\{\mathbf{a}^\loc\right\}_{\loc\in L}$ and $\left\{b^\loc\right\}_{\loc\in L}$ as vector and scalar variables,
respectively, and will use $\mathbf{c}^\loc,d^\loc$ as vector/scalar linear expressions over
$\left\{\mathbf{a}^{\loc'}\right\}_{\loc'\in L}$ and $\left\{b^{\loc'}\right\}_{\loc'\in L}$ to be determined by $\OneStep$ (cf. Item 8 below), respectively.
Similarly, for a transition $\tau$ we will also use
$\mathbf{c}^{\loc,\tau}$ (resp. $\mathbf{c}^{\tau}$), $d^{\loc,\tau}$ (resp., $d^{\tau}$) as
linear expressions over
$\left\{\mathbf{a}^{\loc'}\right\}_{\loc'\in L}$ and $\left\{b^{\loc'}\right\}_{\loc'\in L}$.

\item {\em $\Basic$:} We will use the following notation $\Basic$ for polyhedrons (or
half-spaces) given by
\[
\Basic(\mathbf{c}^\ell,d^\ell,\epsilon)=
\left\{\vec{x}\in\mathbb{R}^{|\pvars|} \mid \left(\mathbf{c}^\ell\right)^\mathrm{T}\cdot\mathbf{x}\le d^\ell-\epsilon\right\};
\]
and similarly, $\Basic(\mathbf{c}^\tau,d^\tau,\epsilon)$ and $\Basic(\mathbf{c}^{\ell,\tau},d^{\ell,\tau},\epsilon)$.

\item {\em $\OneStep$:} We will use $\OneStep(\mathbf{c}^\ell,d^\ell,\epsilon)$ to denote the following predicate:
\[
{\left(\mathbf{c}^\ell\right)}^\mathrm{T}\mathbf{x}\le d^\ell-\epsilon~~\Leftrightarrow~~\mathrm{pre}_\eta(\ell,\mathbf{x})\le\eta(\loc,\mathbf{x})-\epsilon,
\]
and similarly for $\OneStep(\mathbf{c}^{\ell,\tau},d^{\ell,\tau},\epsilon)$.
For $\tau\in\mapsto$ from a non-deterministic location $\loc$ to a target location $\loc'$,
we use $\OneStep(\mathbf{c}^\tau,d^\tau,\epsilon)$ to denote the following predicate:
\[
{\left(\mathbf{c}^\tau\right)}^\mathrm{T}\mathbf{x}\le d^\tau-\epsilon~~\Leftrightarrow~~\eta(\loc',\mathbf{x})\le\eta(\loc,\mathbf{x})-\epsilon~.
\]
By Item 1 in {\bf Algorithm \LRSMSynth} (cf. below), one can observe that $\OneStep$ determines $\mathbf{c}^\ell,d^\ell, \mathbf{c}^{\ell,\tau},d^{\ell,\tau},\mathbf{c}^\tau,d^\tau$ in terms of $\mathbf{a}^\ell,b^\ell$.
\end{compactenum}
\end{notations}

\smallskip\noindent{\em Running example.}
Since our algorithm is technical, we will illustrate the steps of the algorithm
on the running example.
We consider the SGS of Figure~\ref{fig:Q4:SGS}, and
assign the invariant $I$ such that $I(\loc_0)=I(\loc_1)=\mathrm{true}$,  $I(\loc_i)=x\ge 0$ for $2\leq i \leq 8$, $I(\loc_9)=x<0$.

\smallskip\noindent{\bf Algorithm \LRSMSynth.}
Intuitively, our algorithm transform conditions C1-C4 in Definition~\ref{def:lrsm} into Farkas' or Motzkin assertions; the transformation differs among different types of locations.

The steps of algorithm \LRSMSynth\ are as follows:
(i)~the first two steps are related to initialization;
(ii)~then steps~3--5 specify condition C4 of LRSM, where
step~3 considers probabilistic locations, step~4 deterministic locations,
and step~5 both angelic and demonic locations;
(iii)~step~6 specifies condition C2 and step~7 specifies condition C3 of
LRSM;
and (iv)~finally, step~8 integrates all the previous steps into a set of
constraints.
We present the algorithm, and each of the steps~3--7 are
illustrated on the running example immediately after the algorithm.
Formally, the steps are as follows:

\begin{enumerate}\itemsep1pt \parskip0pt \parsep0pt

\item {\em Template.}
The algorithm assigns a template $\eta$ for an LRSM by setting
$\eta(\loc,\mathbf{x}):=(\mathbf{a}^\loc)^\mathrm{T}\mathbf{x}+b^\loc$ for each $\loc$ and
$\mathbf{x}\in\mathbb{R}^{|\pvars|}$.
This ensures the linearity condition C1 (cf. Item 6 in Notations~\ref{notation})

\item {\em Variables for martingale difference and terminating negativity.}
The algorithm assigns a variable $\epsilon$ and variables $K,K'$.

\item {\em Probabilistic locations.}
For each probabilistic location $\loc\in L\backslash\left\{\loc_P^{\lout}\right\}$,
the algorithm transforms the open sentence $\Op(\loc)$
equivalently into
\[
\bigwedge_{k=1}^{k_{\loc}} \left[H_{k,\loc}\subseteq \Basic(\mathbf{c}^\ell,d^\ell,\epsilon) \right]
\]
such that $\OneStep(\mathbf{c}^\ell,d^\ell,\epsilon)$.
Using Farkas' Lemma, Lemma~\ref{lemm:closedness} and Lemma~\ref{lemm:linearemptiness}, the algorithm further transforms it equivalently into the Farkas linear assertion
\[
\phi_\loc:=\bigwedge_{k=1}^{k_{\loc}}\Phi\left[\NS{H_{k,\loc}},\mathbf{c}^\loc,d^\loc-\epsilon\right]\left(\xi^{k,\loc}_\mathrm{p}\right)
\]
where
we have fresh variables $\left\{\xi^{k,\loc}_\mathrm{p}\right\}_{k,\loc}$ (cf. Notations~\ref{notation} for the meaning of $H_{k,\loc}, \OneStep, \Basic, \NS{\centerdot}, \Op$ etc.).

\item {\em Deterministic locations.}
For each deterministic location $\loc\in L\backslash\left\{\loc_P^{\lout}\right\}$,
the algorithm transforms the open sentence $\Op(\loc)$
equivalently into
\[
\bigwedge_{k=1}^{k_{\loc}}\bigwedge_{\tau\in\transitions_\loc}\bigwedge_{\phi'\in G(\tau)} \bigg[H_{k,\loc}\wedge \phi' \subseteq
\Basic(\mathbf{c}^{\ell,\tau},d^{\ell,\tau},\epsilon)
\bigg]
\]
such that the sentence $\forall\mathbf{x}\in G(\tau).\OneStep(\mathbf{c}^{\ell,\tau},d^{\ell,\tau},\epsilon)$ holds.
Using Farkas' Lemma, Lemma~\ref{lemm:closedness} and Lemma~\ref{lemm:linearemptiness}, the algorithm further transforms it equivalently into
\[
\phi_\loc:=\bigwedge_{k=1}^{k_{\loc}}\bigwedge_{\tau\in\transitions_\loc}\bigwedge_{\phi'\in G(\tau)}\Phi\left[\NS{H_{k,\loc}\wedge\phi'},\mathbf{c}^{\loc,\tau},d^{\loc,\tau}-\epsilon\right]\left(\xi^{k,\loc,\phi'}_\mathrm{dt}\right)
\]
where we have fresh variables $\left\{\xi^{k,\loc,\phi'}_\mathrm{dt}\right\}_{k,\loc,\phi'}$.

\item {\em Demonic and angelic locations.}
For each demonic (resp. angelic) location $\loc$, the algorithm transforms the open sentence
$\Op(\loc)$
equivalently into
\[
\displaystyle \bigwedge_{k=1}^{k_{\loc}}\bigg[H_{k,\loc}\subseteq \Operator_{\tran\in\transitions_\loc}
\Basic(\mathbf{c}^{\tran},d^{\tran},\epsilon)
\bigg]
\]
where $\Operator$ is $\bigcap$ for demonic location and $\bigcup$ for angelic location,
such that $\OneStep(\mathbf{c}^{\tran},d^{\tran},\epsilon)$ holds.

\noindent{\em Demonic case.}
Using Farkas' Lemma, Lemma~\ref{lemm:closedness} and Lemma~\ref{lemm:linearemptiness}, the algorithm further transforms the open sentence equivalently
into the linear assertion
\[
\phi_\loc:=\bigwedge_{k=1}^{k_{\loc}}\bigwedge_{\tran\in\transitions_\loc}\Phi\left[\NS{H_{k,\loc}},\mathbf{c}^{\tran},d^{\tran}-\epsilon\right]\left(\xi_\mathrm{dm}^{k,\loc,\tau}\right)
\]
where we have fresh variables $\left\{\xi^{k,\loc,\tau}_\mathrm{dm}\right\}_{1\le k\le k_\loc}$.

\noindent{\em Angelic case.}
The algorithm further transforms the sentence equivalently into
\[
\bigwedge_{k=1}^{k_{\loc}}\bigg[H_{k,\loc}\cap\bigcap_{\tran\in\transitions_\loc}\left\{\mathbf{x}\in\mathbb{R}^{|\pvars|}\mid \left(\mathbf{c}^{\tran}\right)^\mathrm{T}\mathbf{x}> d^{\tran}-\epsilon\right\}=\emptyset\bigg]\enskip;
\]
finally, from Motzkin's Transposition Theorem, Lemma~\ref{lemm:closedness} and Lemma~\ref{lemm:linearemptiness}, the algorithm transforms the sentence equivalently into the nonlinear
constraint (Motzkin assertion)
\[
\phi_{\loc}:=\bigwedge_{k=1}^{k_\loc}\Psi\left[\NS{H_{k,\loc}},\mathbf{C}^\loc,\mathbf{d}^\loc\right]\left(\{\xi^{k,\loc}_{\mathrm{ag}}\}_{k,\loc},\{\zeta^{\tran}_{\mathrm{ag}}\}_{\tran\in\transitions_\loc}\right)
\]
with fresh variables 
$\left\{\xi^{k,\loc}_\mathrm{ag}\right\}_{1\le k\le k_\loc},\left\{\zeta^{\tran}_\mathrm{ag}\right\}_{\tran\in\transitions_\loc}$
where
\[
\mathbf{C}^{\loc}=-\begin{pmatrix}\left(\mathbf{c}^{\tran_1}\right)^{\mathrm{T}}\\\vdots\\{\left(\mathbf{c}^{\tran_m}\right)}^\mathrm{T}\end{pmatrix}\mbox{ and }\mathbf{d}^{\loc}=-\begin{pmatrix}d^{\tran_1} \\\vdots\\d^{\tran_m}\end{pmatrix}+\epsilon\cdot\mathbf{1}
\]
with $\transitions_\loc=\{\tran_1,\dots,\tran_m\}$.

\item {\em Non-negativity for non-terminating location.}
For each location $\loc$ other than the terminating location $\ell_{P}^{\lout}$, the algorithm transforms the open sentence
\[
\varphi_\loc:=\forall\mathbf{x}.\left(\mathbf{x}\in \bigcup I(\loc)\rightarrow\eta(\loc,\mathbf{x})\ge 0\right)
\]
equivalently into
\[
\bigwedge_{k=1}^{k_\loc}\left[H_{k,\loc}\subseteq \Basic(-\mathbf{a}^\loc, b^\loc,0)
\right]\enskip.
\]
Using Farkas' Lemma, Lemma~\ref{lemm:closedness} and Lemma~\ref{lemm:linearemptiness}, the algorithm further transforms it equivalently into
\[
\varphi_\loc:=\bigwedge_{k=1}^{k_\loc}\Phi\left[\NS{H_{k,\loc}}, -\mathbf{a}^\loc,b^\loc\right](\xi^{k,\loc}_{\mathrm{nt}})
\]
where $\xi^{k,\loc}_{\mathrm{nt}}$ are fresh variables.

\item {\em Negativity for terminating location.}
For the terminating location $\ell_{P}^{\lout}$, the algorithm transforms the open sentence
\[
\varphi_{\loc_{P}^{\lout}}:=\forall\mathbf{x}.\left(\mathbf{x}\in \bigcup I(\ell_{P}^{\lout})\rightarrow \left(K'\le\eta(\loc_{P}^{\lout},\mathbf{x})\le K\right)\right)
\]
equivalently into
\begin{align*}
&\bigwedge_{k=1}^{k_{\loc_P^\lout}}\left[H_{k,\loc}\subseteq \Basic(\mathbf{a}^{\loc_P^\lout},-b^{\loc_P^\lout},-K)\right]\\
&\wedge \bigwedge_{k=1}^{k_{\loc_P^\lout}}\left[H_{k,\loc}\subseteq\Basic(-\vec{a}^{\loc_P^\lout},b^{\loc_P^\lout},K')\right]\enskip.
\end{align*}
Using Farkas' Lemma, Lemma~\ref{lemm:closedness} and Lemma~\ref{lemm:linearemptiness}, the algorithm further transforms it equivalently into
\begin{align*}
\varphi_{\ell_{P}^{\lout}}:=&\bigwedge_{k=1}^{k_{\ell_{P}^{\lout}}}\Phi\left[\NS{H_{k,{\ell_{P}^{\lout}}}}, \mathbf{a}^{\loc_P^\lout},-b^{\loc_P^\lout}+K\right]\left(\xi^{k,{\loc_P^\lout}}_{\mathrm{t}}\right)\\
&\wedge\bigwedge_{k=1}^{k_{\ell_{P}^{\lout}}}\Phi\left[\NS{H_{k,{\loc_P^\lout}}}, -\mathbf{a}^{\loc_P^\lout},b^{\loc_P^\lout}-K'\right]\left(\xi^{k,\ell_{P}^{\lout}}_{\mathrm{tt}}\right)
\end{align*}
where $\xi^{k,\loc}_{\mathrm{t}}$'s and $\xi^{k,\loc}_{\mathrm{tt}}$'s are fresh variables.

\item {\em Solving the constraint problem.} For each location $\loc$,
let $\phi_\loc$ and $\varphi_\loc$ be the formula obtained in steps~3--5, and
steps~6--7, respectively.
The algorithm outputs whether the following formula is satisfiable:
\[
\Xi_P:=\epsilon\ge 1\wedge K,K'\le -1\wedge\bigwedge_{\loc\in L}\left(\phi_\loc\wedge \varphi_\loc\right)\enskip,
\]
where the satisfiability is interpreted over all relevant open variables in $\Xi_P$.
\end{enumerate}

\begin{example}{(Illustration of algorithm \LRSMSynth\ on running example).}
We describe the steps of the algorithm on the running example.
For the sake of convenience, we abbreviate $\mathbf{a}^{\loc_i},b^{\loc_i}$ by $a_i,b_i$.
\begin{compactitem}
\item {\em Probabilistic location: step~3.}
In our example,
\[
\phi_{\loc_2}=\Phi[x\ge 0, 0.6a_3+0.4a_4-a_2, b_2-0.6b_3-0.4b_4-\epsilon]
\]
\item {\em Deterministic location: step~4.}
In our example,
\[
\phi_{\loc_0}=\Phi[\mathbb{R},-a_0, b_0-b_1-\epsilon]
\]
\begin{align*}
&\phi_{\loc_1}=\Phi[x\ge 0, a_2-a_1, b_1-b_2-\epsilon]\\
&\qquad\wedge\Phi[x\le 0,a_9-a_1, b_1-b_9-\epsilon]
\end{align*}
\[
\phi_{\loc_i}=\Phi[x\ge 0, a_1-a_i, b_i-b_1-\epsilon-a_1]\mbox{ for }i\in\{5,7\}
\]
\[
\phi_{\loc_i}=\Phi[x\ge 0, a_1-a_i, b_i-b_1-\epsilon+a_1]\mbox{ for }i\in\{6,8\}\enskip.
\]
\item {\em Demonic location: step~5a.}
In the running example,
\begin{align*}
&\phi_{\loc_4}:=\Phi[x\ge 0,a_8-a_4, b_4-b_8-\epsilon]\\
&\qquad\wedge\Phi[x\ge 0, a_7-a_4,b_4-b_7-\epsilon]\enskip.
\end{align*}
\item {\em Angelic location: step~5b.}
In our example,
\[
\phi_{\loc_3}=\Psi\left[x\ge 0,\begin{pmatrix}a_3-a_5\\a_3-a_6\end{pmatrix},\begin{pmatrix}-b_3+b_5+\epsilon\\-b_3+b_6+\epsilon\end{pmatrix}\right]\enskip.
\]
\item {\em Non-negativity of non-terminating location: step~6.}
In our example, we have
\begin{eqnarray*}
&\varphi_{\loc_i}=\Phi[\mathbb{R},-a_i,b_i] \mbox{ for }i\in\{0,1\}\mbox{ and }\\
&\varphi_{\loc_i}=\Phi[x\ge 0,-a_i,b_i] \mbox{ for }2\le i\le 8\enskip.
\end{eqnarray*}

\item {\em Terminating location: step~7.}
In our example, we have
\[
\varphi_{\loc_9}=\Phi[x\le 0,a_9,-b_9+K]\wedge \Phi[x\le 0,-a_9,b_9-K']\enskip.
\]
\end{compactitem}
\end{example}

\begin{remark}\label{rem:thm1}
Note that it is also possible to follow the usage of Motzkin's Transposition Theorem
in~\cite{DBLP:conf/sas/KatoenMMM10} for angelic locations to first turn the formula
into a conjunctive normal form and then apply Motzkin's Theorem on
each disjunctive sub-clause.
Instead we present a direct application of Motzkin's Theorem.
\end{remark}

\smallskip\noindent{\em Correctness and analysis.}
The construction of the algorithm ensures that there exists
an LRSM iff the algorithm \LRSMSynth\ answers yes.
Also note that if \LRSMSynth\ answers yes, then a witness LRSM can be
obtained (for synthesis) from the solution of the constraints.
Moreover, given a witness we obtain an upper bound $\UB(P)$ on $\Eval(P)$
from Theorem~\ref{thm:supermartingale-correctness}.
We now argue two aspects:
\begin{compactenum}
\item {\em Linear constraints.} First observe that for algorithm
\LRSMSynth, all steps, other than the one for angelic non-determinism,
only generates linear constraints.
Hence it follows that in the absence of angelic non-determinism we obtain
a set of linear constraints that is polynomial in the size of the input.
Hence we obtain the second item of Theorem~\ref{thm:mainalgorithm}.

\item {\em Quadratic constraints.}
Finally, observe that for angelic non-determinism the application of Motzkin's
Theorem generates only quadratic constraints. Since the existential first-order theory of the reals can be decided in $\PSPACE$~\cite{Canny:FOreals}, we get the first item of  Theorem~\ref{thm:mainalgorithm}.

\end{compactenum}


\subsection{Lower bound}\label{subsec:hard1}
We establish the third item of Theorem~\ref{thm:mainalgorithm}
(detailed proof in the appendix).

\begin{lemma}\label{lemm:hard1}
The \LRAPP\ realizability problem for \APP s with angelic non-determinism is $\NP$-hard, 
even for non-probabilistic non-demonic programs with simple guards.
\end{lemma}
\begin{proof}[Proof (sketch)]
We show a polynomial reduction from $3$-\textsc{SAT} to the \LRAPP\ realizability problem. 
For a propositional formula $\psi$ we construct a non-probabilistic non-demonic program $P_{\psi}$ 
whose variables correspond to the variables of $\psi$ and whose form is as follows: 
the program consists of a single while loop within which each variable is set to $0$ or $1$ 
via an angelic choice. 
The guard of the loop checks whether $\psi$ is satisfied by the assignment: 
if it is not satisfied, then the program proceeds with another iteration of the loop, 
otherwise it terminates. 
The test can be performed using a propositionally linear predicate; e.g. 
for the formula $(x_1 \vee x_2 \vee \neg x_3) \wedge (\neg x_2 \vee x_3 \vee  x_4)$ 
the loop guard will be $x_1 + x_2 + (1-x_3) \leq \frac{1}{2} \vee (1-x_2) + x_3 + x_4 \leq \frac{1}{2}$.
To each location we assign a simple invariant $I$ which says that all program variables have values between $0$ and $1$. 
The right hand sides of inequalities in the loop guard are set to $\frac{1}{2}$ in order for the reduction to work with this invariant: 
setting them to $1$, which might seem to be an obvious first choice, would only work for an invariant saying that all variables have value $0$ or $1$, 
but such a condition cannot be expressed by a polynomially large propositionally linear predicate.

If $\psi$ is not satisfiable, then the while loop obviously never terminates and hence by Theorem~\ref{thm:supermartingale-correctness} there is no LRSM for $P_{\psi}$ with respect to any invariant, including~$I$. 
Otherwise there is a satisfying assignment $\nu$ for $\psi$ which can be used to construct an LRSM $\mart$ with respect to $I$. 
Intuitively, $\mart$ measures the distance of the current valuation of program variables from the satisfying assignment $\nu$. 
By using a scheduler $\sigma$ that consecutively switches the variables to the values specified by $\nu$ the angel ensures that $\mart$ eventually decreases to zero. Since the definition of a pre-expectation is independent of the scheduler used, we must ensure that the conditions C2 and C4 of LRSM
hold also for those valuations $\vec{x}$ that are not reachable under $\sigma$. 
This is achieved by multiplying the distance of each given variable $x_i$ from $\nu$ by a suitable \emph{penalty factor} $\satpenalty$ in all locations in the loop that are positioned after the branch in which $x_i$ is set. For instance, in a location that follows the choice of $x_1$ and $x_2$ and precedes the choice of $x_3$ and $x_4$ the expression assigned by $\mart$ in the above example will be of the form $\satpenalty\cdot ((1-x_1) + x_2) + (1-x_3) + x_4 + d$, where $d$ is a suitable number varying with program locations. This ensures that the value of $\mart$ for valuations that are not reachable under $\sigma$ is very large, and thus it can be easily decreased in the following steps by switching to $\sigma$. 
\end{proof}

\section{\LRAPP: Quantitative Analysis}\label{sec:quan}
In this section we consider the quantitative questions for \LRAPP.
We first show a program $P$ in \LRAPP{} with only discrete probabilistic
choices such that the expected termination time $\Eval(P)$ is {\em irrational}.

\lstset{language=affprob}
\lstset{tabsize=3}
\newsavebox{\figexampleirrationaltime}
\begin{lrbox}{\figexampleirrationaltime}
\begin{lstlisting}[mathescape]
$n:=1$;
while $n\geq 1$ do
if $\textbf{prob}\mathbf{(}\frac{1}{2}\mathbf{)}$ then $n:=n+1$
    else $n:=n-1$; $n:=n-1$ fi od
\end{lstlisting}
\end{lrbox}
\begin{figure}[h]
\centering
\usebox{\figexampleirrationaltime}
\caption{An example where $\Eval(P)$ is irrational.}
\label{fig:irrational-time}
\end{figure}

\begin{example}\label{ex:irrat2}
Consider the example in Fig.~\ref{fig:irrational-time}. The program $P$ in the figure represents an operation of a so called one-counter Markov chain, a very restricted class of \APP s without non-determinism and with a single integer variable. It follows from results of~\cite{BEKK:pPDA-survey-FMSD} and~\cite{EKM:prob-PDA-expectations} that the termination time of $P$ is equal to a solution of a certain system of quadratic equations, which in this concrete example evaluates to $2(5+\sqrt{5})$, an irrational number (for the precise computation see appendix).
\end{example}

Given that the expected termination time can be irrational, we focus on the problem of its approximation.
To approximate the termination time we first compute \emph{concentration bounds} (see Definition~\ref{def:quanttq}).
Concentration bounds can only be applied if there exist bounds
on martingale change in every step. Hence we define the class of bounded \LRAPP.

\smallskip\noindent{\em Bounded \LRAPP.}
An LRSM $\eta$ wrt invariant $I$ is {\em bounded} if there exists an interval  $[a,b]$
such that the following holds: for all locations $\loc$ and successors $\loc'$ of $\loc$,
and all valuations $\mathbf{x} \in  \bigcup I(\loc)$ and $\mathbf{x}' \in  \bigcup I(\loc')$
if $(\loc',\mathbf{x}')$ is reachable in one-step from $(\loc,\mathbf{x})$, then
we have $(\eta(\loc',\mathbf{x}') - \eta(\loc,\mathbf{x})) \in [a,b]$.
Bounded \LRAPP~ is the subclass of \LRAPP~ for which there exist bounded LRSMs for
some invariant.
For example, for a program $P$, if all updates are bounded by some constants
(e.g., bounded domain variables, and each probability distribution has a bounded range),
then if it belongs to \LRAPP, then it also belongs to bounded \LRAPP.
Note that all examples presented in this section (as well several in
Section~\ref{sec:exp}) are in bounded \LRAPP.

We formally define the quantitative approximation problem for \LRAPP s as follows: the input is
a program $P$ in bounded \LRAPP,
an invariant $I$ for $P$, a bounded LRSM $\eta$ with a bounding interval $[a,b]$, and a rational number $\delta \geq 0$.
%
The output 
is a rational number $\nu$ such that $|\inf_{\sigma \in \martcor{\mart}}\sup_{\pi}\Eval(P)-\nu|\leq \delta$, where $\martcor{\mart}$ is the set of all angelic schedulers $\sigma$ that are \emph{compatible} with $\mart$, i.e. that
obeys the construction for angelic scheduler illustrated below Theorem~\ref{thm:supermartingale-correctness}.
Note that $\martcor{\mart}$ is non-empty (see Remark~\ref{rmk:scheduler}). This condition is somewhat restrictive, as it might happen that no near-optimal angelic scheduler is compatible with a martingale $\eta$ computed via methods in Section~\ref{sec:lrapp}. On the other hand, this definition captures the problem of extracting, from a given LRSM $\eta$, as precise information about the expected termination time as possible. Note that for programs without angelic non-determinism the problem is equivalent to approximating $\Eval(P).$

Our main results on are summarized below. 

\begin{theorem}\label{thm:quan_lrapp}
\begin{enumerate}
\item
A concentration bound $B$ can be computed in the same complexity as for
qualitative analysis (i.e., in polynomial time with only demonic non-determinism,
and in $\PSPACE$ in the general case).
Moreover, the bound $B$ is at most exponential.
\item
The quantitative approximation problem can be solved in a doubly exponential time for bounded \LRAPP{} with only discrete probability choices. It cannot be solved in polynomial time unless $\mathsf{P} = \PSPACE$, even for programs without probability or non-determinism.
\end{enumerate}
\end{theorem}

\begin{remark}
Note that the bound $B$ is exponential, and our result
(Lemma~\ref{lemm:hard2}) shows that there exist deterministic programs in
bounded \LRAPP{} that terminate
exactly after exponential number of steps (i.e., an exponential bound for
$B$ is asymptotically optimal for bounded \LRAPP).
\end{remark}

\subsection{Concentration Results on Termination Time}\label{subsec:concentration}
In this section, we present the first approach to show how LRSMs can be used to
obtain concentration results on termination time for bounded \LRAPP.

\subsubsection{Concentration Inequalities}
We first consider Azuma's Inequality~\cite{Azuma1967inequality} which serves as a basic concentration inequality on
supermartingales, and then adapt finer inequalities such as Hoeffding's Inequality~\cite{Hoeffding1963inequality,ColinMcDiarmid1998concentration},
and Bernstein's Inequality~\cite{Berstein1962inequality,ColinMcDiarmid1998concentration} to supermartingales.

\begin{theorem}[Azuma's Inequality~\cite{Azuma1967inequality}]
Let $\{X_n\}_{n\in\mathbb{N}}$ be a supermartingale wrt some filtration  $\{\mathcal{F}_n\}_{n\in\mathbb{N}}$ and $\{c_n\}_{n\in\mathbb{N}}$ be a sequence of positive numbers. If $|X_{n+1}-X_n|\le c_n$ for all $n\in\mathbb{N}$, then
\[
\mathbb{P}(X_n-X_1\ge \lambda)\le e^{-\frac{\lambda^2}{2\cdot\sum_{k=2}^n c_n^2}}
\]
for all $n\in\mathbb{N}$ and $\lambda>0$.
\end{theorem}

Intuitively, Azuma's Inequality bounds the amount of actual increase of a
supermartingale at a specific time point.
It can be refined by Hoeffding's Inequality.
The original Hoeffding's Inequality~\cite{Hoeffding1963inequality} works for
martingales; and we show how to extend to
supermartingales.

\begin{theorem}[Hoeffding's Inequality on Supermartingales]\label{thm:hoeffding}
Let $\{X_n\}_{n\in\mathbb{N}}$ be a supermartingale wrt some filtration $\{\mathcal{F}_n\}_{n\in\mathbb{N}}$ and $\{[a_n,b_n]\}_{n\in\mathbb{N}}$ be a sequence of intervals of positive length in $\mathbb{R}$.
If $X_1$ is a constant random variable and $X_{n+1}-X_n\in [a_n,b_n]$ a.s. for all $n\in\mathbb{N}$, then
\[
\mathbb{P}(X_n-X_1\ge\lambda)\le e^{-\frac{2\lambda^2}{\sum_{k=2}^n(b_k-a_k)^2}}
\]
for all $n\in\mathbb{N}$ and $\lambda> 0$.
\end{theorem}

\begin{remark}
By letting the interval $[a,b]$ be $[-c,c]$ in Hoeffding's inequality, we obtain Azuma's inequality.
Thus, Hoeffding's inequality is at least as tight as Azuma's inequality and is strictly tighter when $[a,b]$ is not a symmetric interval.
\end{remark}

If variation and expected value of differences of a supermartingale is considered, then
Bernstein's Inequality yields finer concentration than Hoeffding's Inequality.

\begin{theorem}[Bernstein's Inequality~\cite{Berstein1962inequality,ChungConcentrationSurvey}]\label{thm:bernstein}
Let $\{X_n\}_{n\in\mathbb{N}}$ be a supermartingale wrt some filtration $\{\mathcal{F}_n\}_{n\in\mathbb{N}}$ and $M\ge 0$.
If $X_1$ is constant, $X_{n+1}-\mathbb{E}(X_{n+1}\mid\mathcal{F}_{n})\le M$ a.s. and
$\mathrm{Var}(X_{n+1}\mid \mathcal{F}_{n})\le c^2$ for all $n\ge 1$, then
\[
\mathbb{P}(X_n-X_1\ge\lambda)\le e^{-\frac{\lambda^2}{2(n-1)c^2+\frac{2M\lambda}{3}}}
\]
for all $n\in\mathbb{N}$ and $\lambda>0$.
\end{theorem}

\subsubsection{LRSMs for Concentration Results}

The only previous work which considers concentration results for probabilistic
programs is~\cite{SriramCAV}, that argues that Azuma's Inequality can be used to
obtain bounds on deviations of program variables.
However, this technique does not present concentration result on termination time.
For example, consider that we have an additional program variable to measure the number
of steps.
But still the invariant $I$ (wrt which the LRSM is constructed) can ignore the additional
variable, and thus the LRSM constructed need not provide information about termination time.
We show how to overcome this conceptual difficulty.
For the rest of this section, we fix a program $P$ in bounded \LRAPP{} and
its SGS $\mathcal{G}_P$.
We first present our result for Hoeffding's Inequality (for bounded
\LRAPP) and then the result for Bernstein's Inequality (for a subclass of
bounded \LRAPP).

For the rest of this part we fix an affine probabilistic program $P$
and let $\mathcal{G}_P=(\locs,(\pvars,\rvars),\ell_0,\vec{x}_0,\transitions,\probdist,\guards)$
be its associated SGS.
We fix the filtration $\{\mathcal{F}_n\}_{n\in\mathbb{N}}$ such that each $\mathcal{F}_n$ is the smallest $\sigma$-algebra on runs that
makes all random variables in $\{\theta_j\}_{1\le j\le n},\{\overline{x}_{k,j}\}_{1\le k\le |\pvars|, 1\le j\le n}$ measurable,
where we recall that $\theta_j$ is the random variable representing the location at the $j$-th step,
and $\overline{x}_{k,j}$ is the random variable representing the value of the program variable $x_{k}$ at the $j$-th step.
We recall that $T$ is the termination-time random variable for $P$.

\smallskip\noindent{\bf Constraints for LRSMs to apply Hoeffding's Inequality.}
Let $\eta$ be an LRSM to be synthesized for $P$ wrt linear invariant $I$.
Let $\{X_n\}_{n\in\mathbb{N}}$ be the stochastic process defined by
\[
X_n:=\eta(\theta_n, \{\overline{x}_{k,n}\}_k)
\]
for all natural numbers $n$.
To apply Hoeffding's Inequality, we need to synthesize constants $a,b$ such that
$X_{n+1}-X_{n}\in [a,b]$ a.s. for all natural numbers $n$.
We encode this condition as follows:
\begin{compactitem}
\item {\em Probablistic or demonic locations.}
for all $\loc\in L_P\cup L_D$ with successor locations $\loc_1,\loc_2$, the following sentence
holds:
\[
\forall\vec{x}\in \bigcup I(\loc). \ \forall i\in\{1,2\}. \ \ \big(\eta(\loc_i,\vec{x})-\eta(\loc,\vec{x})\in [a,b]\big).
\]
\item {\em Deterministic locations.} for all $\loc\in L_S$ and all $\tran=(\loc,f,\loc')\in \mapsto_\loc$,
the following sentence holds:
\[
\forall\vec{x}\in \bigcup I(\loc)\wedge G(\tran). \ \forall\vec{r}. \ \big(\eta(\loc',f(\mathbf{x},\mathbf{r}))-\eta(\loc,\mathbf{x})\in [a,b]\big)
\]
\item {\em Angelic location.} for all $\loc\in L_A$ with successor locations $\loc_1,\loc_2$, the following condition holds:
\[
\forall\vec{x}\in\bigcup I(\loc).\exists i\in\{1,2\}.
\big(a\le \eta(\loc_i,\mathbf{x})-\eta(\loc,\mathbf{x})\le -\epsilon\le b\big)
\]
\item We require that $-\epsilon\in [a,b]$. This is not restrictive since $\epsilon$ reflects the supermartingale difference.
\end{compactitem}
We have that if the previous conditions hold, then $a,b$ are valid constants.
Note that all the conditions above can be transformed into an
existential formula on parameters of $\eta$ and $a,b$ by Farkas' Lemma or
Motzkin's Transposition Theorem, similar to the transformation in
\LRSMSynth\ in Section~\ref{subsec:algo}.
Moreover for bounded \LRAPP{} by definition there exist valid constants
$a$ and $b$.

\smallskip\noindent{\em Key supermartingale construction.}
We now show that given the LRSM $\eta$ and the constants $a,b$ synthesized
wrt the conditions for $\eta$ above,
how to obtain concentration results on termination time.
Define the stochastic process $\{Y_n\}_{n\in\mathbb{N}}$ by:
\[
Y_n=X_n+\epsilon\cdot(\min\{T,n\}-1)\enskip.
\]
The following proposition shows that $\{Y_n\}_{n\in\mathbb{N}}$ is a supermartingale and satisfies the requirements of Hoeffding's Inequaltiy.

\begin{proposition}\label{prop:hoeffding}
$\{Y_n\}_{n\in\mathbb{N}}$ is a supermartingale and $Y_{n+1}-Y_n\in [a+\epsilon,b+\epsilon]$ almost surely for all $n \in \Nats$.
\end{proposition}

\smallskip\noindent{\bf LRSM and supermartingale to concentration result.}
We now show how to use the LRSM and the supermartingale $Y_n$ to achieve the
concentration result.
Let $W_0:=Y_1=(\mathbf{a}^{\loc_P^\lin})^\mathrm{T}\initval+b^{\loc_P^\lin}$.
Fix an angelic strategy that fulfills supermartingale difference and bounded change for LRSM;
and fix any demonic strategy.
By Hoeffing's Inequality, for all $\lambda>0$, we have
$\mathbb{P}(Y_n-W_0\ge\lambda)\le e^{-\frac{2\lambda^2}{(n-1)(b-a)^2}}$.
Note that $T> n$ iff $X_n\ge 0$ by conditions C2 and C3 of LRSM.
Let $\alpha=\epsilon(n-1)-W_0$ and $\wh{\alpha}=\epsilon(\min\{n,T\}-1)-W_0$.
Note that with the conjunct $T>n$ we have that $\alpha$ and $\wh{\alpha}$
coincide.
Thus, for $\mathbb{P}(T > n)=\mathbb{P}(X_n\ge 0\wedge T>n)$ we have
\begin{eqnarray*}
\mathbb{P}(X_n\ge 0\wedge T>n) 
&=&\mathbb{P}((X_n+\alpha \ge \alpha ) \wedge (T>n))\\
&=&\mathbb{P}((X_n+\wh{\alpha} \ge \alpha ) \wedge (T>n))\\
&\le &\mathbb{P}((X_n+ \wh{\alpha} \geq \alpha))\\
&=&\mathbb{P}(Y_n-Y_1\ge \epsilon(n-1)-W_0)\\
&\le &e^{-\frac{2(\epsilon(n-1)-W_0)^2}{(n-1)(b-a)^2}}
\end{eqnarray*}
for all $n>\frac{W_0}{\epsilon}+1$.
The first equality is obtained by simply adding $\alpha$ on both
sides, and the second equality uses that because of the conjunct $T>n$
we have $\min\{n,T\}=n$ which ensures $\alpha=\wh{\alpha}$.
The first inequality is obtained by simply dropping the conjunct $T>n$.
The following equality is by definition, and the final inequality is
Hoeffding's Inequality.
Note that in the exponential function the numerator is quadratic in $n$
and denominator is linear in $n$, and hence the overall function is
exponentially decreasing in $n$.

\smallskip\noindent{\em Computational results for concentration inequality.}
We have the following results which establish the second item of Theorem~\ref{thm:quan_lrapp}:
\begin{compactitem}
\item {\em Computation.} Through the synthesis of the LRSM $\eta$ and $a,b$,
a concentration bound $B_0=\frac{W_0}{\epsilon}+2$ can be computed in $\PSPACE$
in general and in PTIME without angelic nondeterminism (similar
to \LRSMSynth\ algorithm).
\item {\em Optimization.} In order to obtain a better concentration bound $B$,
a binary search can be performed on the interval $[0,B_0]$
to find an optimal $B\in [0,B_0]$ such that $\epsilon\cdot (B-2)\ge W_0$
is consistent with the constraints for synthesis of $\eta$ and $a,b$.

\item {\em Bound on $B$.} Note that since $B$ is computed in polynomial
space, it follows that $B$ is at most exponential.
\end{compactitem}

\begin{remark}[Upper bound on $\Thr(P,x)$]\label{rem:upperbound}
We now show that our technique along with the concentration result
also presents an upper bound for $\Thr(P,x)$ as follows:
To obtain an upper bound for a given $x$, we first search for a large number $M_0$ such that
$M_0({b-a})\le \epsilon(x-1)-W_0$ and $(x-2)\epsilon\ge W_0$ are not consistent with the conditions
for $\eta$ and $a,b$; then we perform a binary search for an $M\in [0,M_0]$ such that the (linear) conditions
$M({b-a})\le \epsilon(x-1)-W_0$ and $(x-2)\epsilon\ge W_0$ are consistent with the conditions
for $\eta$ and $a,b$.
Then $\Thr(P,x) \leq e^{-2M^2/(x-1)}$.
Note that we already provide an upper bound $\UB(P)$ for $\Eval(P)$
(recall Theorem~\ref{thm:supermartingale-correctness} and
Remark~\ref{rem:thm1}), and the upper bound $\UB(P)$ holds for
\LRAPP, not only for bounded \LRAPP.
\end{remark}

\smallskip\noindent{\bf Applying Bernstein's Inequality.}
To apply Bernstein's Inequality, the variance on the supermartingale difference needs to be evaluated,
which might not exist in general for \LRAPP s.
We consider a subclass of \LRAPP s, namely, {\em incremental \LRAPP s}.

\begin{definition}\label{de:inc}
A program $P$ in \LRAPP \ is \emph{incremental} if all variable updates are of the form
$x:=x+g(\mathbf{r})$ where $g$ is some linear function on random variables $\mathbf{r}$.
An LRSM is \emph{incremental} if it has the same coefficients for each program variable at every location, i.e. $\vec{a}^\loc=\vec{a}^{\loc'}$ for all $\loc,\loc'\in L$.
\end{definition}

\begin{remark}
The incremental condition for LRSMs can be encoded as a linear assertion.
\end{remark}

\smallskip\noindent{\em Result.} We show that for incremental \LRAPP,
Bernstein's Inequality can be applied for concentration results on termination
time, using the same technique we developed for applying Hoeffding's Inequality.
The technical details are presented in the appendix.

\subsection{Complexity of quantitative approximation}\label{subsec:hard2}

We now show the second item of Theorem~\ref{thm:quan_lrapp}. 
The doubly exponential upper bound is obtained by obtaining a concentration bound $B$ via aforementioned methods and unfolding the program up to $\mathcal{O}(B)$ steps (details in the appendix).

\begin{lemma}\label{coro:concen}
The quantitative approximation problem can be solved in a doubly exponential time for programs with only discrete probability choices.
\end{lemma}

For the $\PSPACE$ lower bound we use the following lemma.

\begin{lemma}\label{lemm:hard2}
For every $C\in\Nset$ the following problem is $\PSPACE$-hard:
Given a program $P$ without probability or non-determinism, with simple guards, and belonging to bounded \LRAPP;
and a number $N \in \Nset$ such that either
$\Eval(P)\leq N$ or $\Eval(P)\geq N\cdot C$, decide, which of these two alternatives hold.
\end{lemma}
\begin{proof}[Proof (Sketch)]
We first sketch the proof of item 1. Fix a number $C$.
We show a polynomial reduction from the following problem that is $\PSPACE$-hard
for a suitable constant $K$: Given a deterministic Turing machine (DTM) $\mathcal{T}$
such that on every input of length $n$ the machine $\mathcal{T}$ uses at most $K\cdot n$ tape cells,
and given a word $w$ over the input alphabet of $\mathcal{T}$, decide, whether $\mathcal{T}$ accepts $w$.

For a given DTM $\mathcal{T}$ and word $w$ we construct a program $P$ that emulates, through updates of its variables, the computation of $\mathcal{T}$ on $w$. This is possible due to the bounded space complexity of $\mathcal{T}$. The program $P$ consists of a single while-loop whose every iteration corresponds to a single computational step of $\mathcal{T}$. The loop is guarded by an expression $m\geq 1 \wedge r \geq 1$, where $m,r$ are special variables such that $r$ is initialized to 1 and $m$ to $C\cdot J$, where $J$ is such a number that if $\mathcal{T}$ accepts $w$, it does so in at most $J$ steps ($J$ can be computed in polynomial time again due to bounded space complexity of $\mathcal{T}$). The variable $m$ is decremented in every iteration of the loop, which guarantees eventual termination. If it happens during the loop's execution that $\mathcal{T}$ (simulated by $P$) enters an accepting state, then $r$ is immediately set to zero, making $P$ terminate immediately after the current iteration of the loop. Now $P$ can be constructed in such a way that each iteration of the loop takes the same amount $W$ of time. If $\mathcal{T}$ \emph{does not} accept $w$, then $P$ terminates in exactly $C\cdot J\cdot W$ steps. On the other hand, if $\mathcal{T}$ \emph{does} accept $w$, then the program terminates in at most $J\cdot W$ steps. Putting $N= J\cdot W$, we get the proof of the first item.
\end{proof}

\begin{remark}
Both in proofs of Lemma~\ref{lemm:hard1} and Lemma~\ref{lemm:hard2} we have
only variables whose change in each step is bounded by~1.
Hence both the hardness proofs apply to bounded \LRAPP.
\end{remark}

\begin{corollary}
The quantitative approximation problem cannot be solved in polynomial time unless $\mathsf{P}=\mathsf{PSPACE}$. Moreover, $\Eval(P)$ cannot be approximated up to any fixed additive or multiplicative error in polynomial time unless $\mathsf{P}=\mathsf{PSPACE}$.
\end{corollary}

\section{Experimental Results}\label{sec:exp}
In this section we present our experimental results.
First observe that one of the key features of our algorithm \LRSMSynth\
is that it uses only operations that are {\em standard} (such as linear
invariant generation, applying Farkas' Lemma), and have been extensively
used in programming languages as well as in several tools.
Thus the efficiency of our approach is similar to the existing methods
with such operations, e.g.,~\cite{SriramCAV}.
The purpose of this section is to demonstrate the {\em effectiveness}
of our approach, i.e., to show that our approach can answer questions
for which no previous automated methods exist.
In this respect we show that our approach can (i)~handle probabilities
and demonic non-determinism together, and (ii)~provide useful answers for
quantitative questions, and the existing tools do not handle either of them.
By useful answers we mean that the concentration bound $B$ and the upper
bound $\UB(P)$ we compute provide reasonable answers.
To demonstrate the effectiveness we consider several classic examples and
show how our method provides an effective automated approach to reason about
them.
Our examples are (i)~random walk in one dimension;
(ii)~adversarial random walk in one dimension; and
(iii)~adversarial random walk in two dimensions.

\smallskip\noindent{\bf Random walk (RW) in one dimension (1D).}
We consider two variants of random walk (RW) in one dimension (1D).
Consider a RW on the positive reals such that at each time step,
the walk moves left (decreases value) or right (increases value).
The probability to move left is $0.7$ and the probability to move
right is $0.3$.
In the first variant, namely {\em integer-valued} RW, every change in the value
is by~1; and in the second variant, namely {\em real-valued} RW,
every change is according to a uniform distribution in $[0,1]$.
The walk starts at value $n$, and terminates if value zero or less is reached.
Then the random walk terminates almost-surely, however, similar
to Example~\ref{ex:irrat2} even in the integer-valued case the expected
termination time is irrational.

\begin{table}
{\scriptsize
\begin{center}
\begin{tabular}{ |c|c|c|c|c| }
\hline
   &    Time &  $B$  &  $\UB(P)$  &  Init. Cofig. \\
\hline
\multirow{5}{*}{Int RW 1D.} & \multirow{5}{*}{$\le 0.02$ sec.} & 47.00 & 46.00 & 5 \\
\cline{3-5}
& & 84.50 & 83.50 & 10 \\
\cline{3-5}
& & 122.00 & 121.00 & 15 \\
\cline{3-5}
& & 159.50 & 158.50 & 20\\
\cline{3-5}
& & 197.00 & 196.00 & 25 \\
\hline
\multirow{5}{*}{Real RW 1D.} & \multirow{5}{*}{$\le 0.01$ sec.} & 92.00 & 91.00 & 5 \\
\cline{3-5}
& & 167.00 & 166.00 & 10 \\
\cline{3-5}
& & 242.00 & 241.00 & 15 \\
\cline{3-5}
& & 317.00 & 316.00 & 20 \\
\cline{3-5}
& & 392.00 & 391.00 & 25 \\
\hline
\multirow{5}{*}{Adv RW 1D.} & \multirow{5}{*}{$\le 0.01$ sec. } & 41.00 & 40.00 & 5 \\
\cline{3-5}
& & 74.33 & 73.33 & 10 \\
\cline{3-5}
& & 107.67 & 106.67 & 15 \\
\cline{3-5}
& & 141.00 & 140.00 & 20 \\
\cline{3-5}
& & 174.33 & 173.33 & 25 \\
\hline
\multirow{5}{*}{Adv RW 2D.} & \multirow{5}{*}{$\le 0.02$ sec.} & - & 122.00 & (5,10) \\
\cline{3-5}
& & - & 152.00 & (10,10) \\
\cline{3-5}
& & - & 182.00 & (15,10) \\
\cline{3-5}
& & - & 212.00 & (20,10) \\
\cline{3-5}
& & - & 242.00 & (25,10) \\
\hline
\multirow{5}{*}{Adv RW 2D. (Variant)} & \multirow{5}{*}{$\le 0.02$ sec.} & 162.00 & 161.00 & (5,0) \\
\cline{3-5}
& & 262.00 & 261.00 & (10,0) \\
\cline{3-5}
& & 362.00 & 361.00 & (15,0)\\
\cline{3-5}
& & 462.00 & 461.00 & (20,0) \\
\cline{3-5}
& & 562.00 & 561.00 & (25,0)\\
\hline
\end{tabular}
\end{center}
\caption{Experimental results: the first column is the example name, the second
column is the time to solve the problem, the following columns are our concentration bound and
upper bound on termination time  for a given initial condition.
}\label{tab:exp}
}
\end{table}

\smallskip\noindent{\bf Adversarial RW in 1D.}
We consider adversarial RW in 1D that models a discrete queuing system that
perpetually processes tasks incoming from its environment at a known
average rate.
In every iteration there are $r$ new incoming tasks, where $r$ is a random
variable taking value $0$ with probability $\frac{1}{2}$, value $1$ with
probability $\frac{1}{4}$ and value $2$ with probability $\frac{1}{4}$.
Then a task at the head of the queue is processed in a way determined by a
type of the task, which is not known a priori and thus is assumed to be
selected demonically. If an \emph{urgent} task is encountered,
the system solves the task rapidly, in one step, but there is a $\frac{1}{8}$ chance
that this rapid process ends in failure that produces a new task to be handled.
A \emph{standard} task is processed at more leisurely pace, in two steps,
but is guaranteed to succeed.
We are interested whether for any initial number of tasks in the queue the program
eventually terminates (queue stability) and in bounds on expected termination
time (efficiency of task processing).

\smallskip\noindent{\bf Adversarial RW in 2D.}
We consider two variants of adversarial RW in 2D.
\begin{compactenum}
\item {\em Demonic RW in 2D.}
We consider a RW in two dimensions, where at every time step either the
$x$-axis or the $y$-axis changes, according to a uniform distribution in
$[-2,1]$.
However, at each step, the adversary decides whether it is the $x$-axis or
the $y$-axis.
The RW starts at a point $(n_1,n_2)$, and terminates if either the $x$-axis
or $y$-axis is reached.

\item {\em Variant RW in 2D.}
We consider a variant of RW in 2D as follows.
There are two choices: in the first (resp. second) choice
(i)~with probability $0.7$ the $x$-axis (resp. $y$-axis) is
incremented by uniform distribution $[-2,1]$ (resp., $[2,-1]$),
and (ii)~with probability $0.3$, the $y$-axis (resp. $x$ axis)
is incremented by $[-2,1]$ (resp., $[2,-1]$).
In other words, in the first choice the probability to move
down or left is higher than the probability to move up or right;
and conversely in the second choice.
At every step the demonic choice decides among the two choices.
The walk starts at $(n_1,n_2)$ such that $n_1 > n_2$, and
terminates if the $x$-axis value is at most the $y$-axis value
(i.e., terminates for values $(n,n')$ s.t. $n \leq n'$).

\end{compactenum}

\begin{figure}
\centering
\includegraphics[scale=0.5]{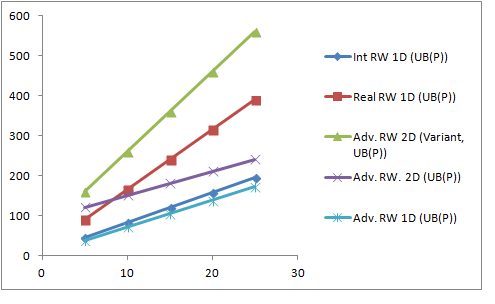}
\caption{The plot of $\UB(P)$ vs the initial location.}
\label{fig:exp}
\end{figure}

\smallskip\noindent{\em Experimental results.}
\label{sec:experiments}
Our experimental results are shown in Table~\ref{tab:exp} and
Figure~\ref{fig:exp}.
Note that all examples considered, other than demonic RW in 2D, are in 
bounded \LRAPP{} (with no non-determinism or demonic non-determinism)
for which all our results are polynomial time.
For the demonic RW in 2D, which is not a bounded \LRAPP\
(for explanation why this is not a bounded \LRAPP\ see
Section~\ref{sec:app_exp} of the appendix),
concentration results cannot be obtained, however we obtain the upper bound
$\UB(P)$ from our results as the example belongs to \LRAPP.
Our experimental result show that the concentration bound and upper bound on
expected termination time (recall $\UB(P)$ from Remark~\ref{rem:thm1})
we compute is a {\em linear} function in all cases (see Fig~\ref{fig:exp}).
This shows that our automated method can effectively compute, or some of the most classical
random walks studied in probability theory, concentration
bounds which are asymptotically tight (the expected number of steps to decrease the value of a standard asymmetric random walk by $n$ is equal to $n$ times the expected number of steps needed to decrease it by $1$, i.e. it is linear in $n$).
For our experimental results, the linear constraints generated by \LRSMSynth\ was
solved by CPLEX~\cite{cplex}.
The programs with the linear invariants are presented in
Section~\ref{sec:app_exp}
of the appendix.

\smallskip\noindent{\em Significance of our result.}
We now highlight the significance of our approach.
The analysis of RW in 1D (even without adversary) is a
classic problem in probability theory, and the expected termination time
can be irrational and involve solving complicated equations.
Instead our experimental results show that using our approach (which is
polynomial time) we can compute upper bound on the expected time that is a
linear function.
This shows that we provide a practical and computational approach for
quantitative reasoning of probabilistic processes.
Moreover, our approach also extends to more complicated probabilistic
processes (such as RW with adversary, as well as in 2D),
and compute upper bounds which are linear, whereas precise mathematical
analysis of such processes is extremely complicated.

\section{Related Work}
We have already discussed several related works,
such as~\cite{MM04,MM05,SriramCAV,BG05,HolgerPOPL} in
Section~\ref{sec:introduction} (Previous results).
We discuss other relevant works here.
The termination for concurrent probabilistic programs under fairness was considered in~\cite{SPH84}.
A sound and
complete characterization of almost-sure termination for countable state space
was given in~\cite{HS85}.
A sound and complete method for proving termination of finite state programs
was given in~\cite{EGK12}.
Termination analysis of non-probabilistic programs has received a lot of
attention over the last decade as
well~\cite{DBLP:conf/cav/BradleyMS05,DBLP:conf/tacas/ColonS01,DBLP:conf/vmcai/PodelskiR04,DBLP:conf/pods/SohnG91,BMS05b,CSZ13,LJB01}.
The most closely related works to our work are~\cite{SriramCAV,BG05,HolgerPOPL}
that consider termination of probabilistic programs via ranking Lyapunov functions and supermartingales.
However, most of the previous works focus on proving a.s. termination and
finite termination, and discuss soundness and completeness.
In contrast, in this work we consider simple (linear) ranking supermartingales,
and study the related algorithmic and complexity issues.
Moreover, we present answers to the quantitative termination questions, and
also consider two types of non-determinism together that has not been considered
before.

\section{Conclusion and Future Work}
In this work we considered the basic algorithmic problems related to
qualitative and quantitative questions for termination of probabilistic
programs.
Since our focus was algorithmic we considered simple (linear) ranking
supermartingales, and established several complexity results.
The most prominent are that for programs with demonic non-determinism the
qualitative problems can be solved in polynomial time, whereas for angelic
non-determinism with no probability the qualitative problems are $\NP$-hard.
We also present $\PSPACE$-hardness results for the quantitative problems,
and present the first method through linear ranking supermartingales to obtain concentration results on termination
time.
There are several directions for future work.
The first direction is to consider special cases of non-linear ranking supermartingales
and study whether efficient algorithmic approaches can be developed for them.
The second interesting direction would be to use the methods of martingale theory to infer deeper insights into the behaviour of probabilistic programs, e.g. via synthesizing assertions about the distribution of program variables ("stochastic invariants").

\bibliographystyle{abbrvnat}
\bibliography{bib_PL}

\clearpage

\appendix

\section{Proofs for Section~\ref{sec:lrapp}}

\noindent{\textbf{Proposition~\ref{prop:rsm}}.
Let $\{X_n\}_{n\in\mathbb{N}}$ be an RSM wrt filtration $\{\mathcal{F}_n\}_{n\in\mathbb{N}}$ and constants $K,\epsilon$ (cf. Definition~\ref{def:rsm}).
Let $Z$ be the random variable defined as
$Z:=\min\{n\in\mathbb{N}\mid X_n<0\}$;
which denotes the first time $n$ that the RSM $X_n$ drops below~0.
Then $\probm(Z<\infty)=1$ and $\expv(Z)\le\frac{\expv(X_1)-K}{\epsilon}$.

\begin{proof}
The proof is similar to~\cite[Lemma 5.5]{HolgerPOPL}.
We first prove by induction on $n\ge 1$ that
\[
\expv(X_n)\le \expv(X_1)-\epsilon\cdot\sum_{k=1}^{n-1}\probm(X_k\ge 0)\enskip.
\]
The base step $n=1$ is clear.
The inductive step can be carried out as follows:
\[
\begin{array}{rcl}
\expv(X_{n+1})&= &\expv\left(\expv(X_{n+1}\mid \mathcal{F}_n)\right) \\[1ex]
&\le &\expv(X_n)-\epsilon\cdot\expv(\mathbf{1}_{X_n\ge 0}) \\[1ex]
&\le &\expv(X_1)-\epsilon\cdot\sum_{k=1}^{n-1}\probm(X_k\ge 0)-\epsilon\cdot\probm(X_n\ge 0) \\[1ex]
&= & \expv(X_1)-\epsilon\cdot\sum_{k=1}^{n}\probm(X_k\ge 0)\enskip.
\end{array}
\]
The first equality is the {\em total expectation law} for conditional
expectation;
the first inequality is obtained from the fact that $X_n$ is a supermartingale;
the second inequality is obtained from the inductive hypothesis;
and the final equality is simply rearranging terms.
Since $X_n\ge K$ almost surely for all $n\in\mathbb{N}$, we have that
$\expv(X_{n}) \geq K$, for all $n$.
Hence from above we have that
\[
\sum_{k=1}^{n}\probm(X_k\ge 0) \leq \frac{\expv(X_1) - \expv(X_{n+1})}{\epsilon} \leq \frac{\expv(X_1) -K}{\epsilon}.
\]
Hence the series $\sum_{k=1}^{\infty}\probm(X_k\ge 0)$ converges and
\[
\sum_{k=1}^{\infty}\probm(X_k\ge 0)\le \frac{\expv(X_1)-K}{\epsilon}\enskip.
\]
It follows from $Z\ge k\Rightarrow X_k\ge 0$ that
\begin{itemize}
\item $\probm(Z=\infty)=\lim_{k\rightarrow\infty}\probm(Z\ge k)=0$ and
\item $\expv(Z)\le\sum_{k=1}^{\infty}\probm(X_k\ge 0)\le \frac{\expv(X_1)-K}{\epsilon}$.
\end{itemize}
The desired result follows.
\end{proof}

To prove Theorem~\ref{thm:supermartingale-correctness}, we need the following lemma which specifies
the relationship between pre-expectation and conditional expectation.

\begin{lemma}\label{lemma:condexp}
Let $\eta$ be an LRSM and $\sigma$ be the angelic scheduler whose decisions optimize the value of $\eta$
at the last configuration of any finite path, represented by
\[
\sigma(\loc,\vec{x})=\mathrm{argmin}_{(\loc,f,\loc')\in\transitions}\eta(\loc',\vec{x})
\]
for all end configuration $(\loc,\vec{x})$ such that
$\loc\in L_A$ and $\vec{x}\in\mathbb{R}^{|\pvars|}$.
Let $\pi$ be any demonic scheduler.
Let the stochastic process $\{X_n\}_{n\in\mathbb{N}}$ be defined such that
\[
X_n:=\eta(\theta_n,\{\overline{x}_{k,n}\}_{1\le k\le|\pvars|})\enskip.
\]
Then for all $n\in\mathbb{N}$,
\[
\expv^{\sigma,\pi}(X_{n+1}\mid\mathcal{F}_n)\le\mathrm{pre}_\eta(\theta_n,\{\overline{x}_{k,n}\}_{1\le k\le|\pvars|})\enskip.
\]
\end{lemma}

\begin{proof}
For all $n\in\mathbb{N}$, from the program syntax we have
\[
X_{n+1}= \mathbf{1}_{\theta_n=\loc_p^\lout}\cdot X_n+Y_P+Y_S+Y_A+Y_D
\]
where the terms are described below:
\[
Y_P:=\sum_{\loc\in L_P}\Big[\mathbf{1}_{\theta_n=\loc}\cdot \sum_{i\in\{0,1\}} \big(\mathbf{1}_{B_\loc=i}\cdot\eta(\loc_{B_\loc=i},\{\overline{x}_{k,n}\}_{k}) \Big]
\]
where each random variable $B_\loc$ is the Bernoulli random variable for the decision of the probabilistic branch and $\loc_{B_\loc=0},\loc_{B_\loc=1}$ are the corresponding successor locations of $\loc$.
Note that all $B_\loc$'s and $\mathbf{r}$'s are independent of $\mathcal{F}_n$.
In other words, $Y_P$ describes the semantics of probabilistic locations.
\[
Y_S:=\sum_{\loc\in L_S}\sum_{(\loc,f,\loc')\in\mapsto_\loc}\Big[\mathbf{1}_{\theta_n=\loc\wedge \{\overline{x}_{k,n}\}_{k}\in G(\tran)}\cdot \eta(\loc',f(\{\overline{x}_{k,n}\}_{k},\mathbf{r})\Big];
\]
describes the semantics of deterministic locations.
\[
Y_A:=\sum_{\loc\in L_A}\mathbf{1}_{\theta_n=\loc}\cdot\eta(\sigma(\loc,\{\overline{x}_{k,n}\}_k),\{\overline{x}_{k,n}\}_k);
\]
describes the semantics of angelic locations, where $\sigma(\loc,\{\overline{x}_{k,n}\}_k)$ here denotes the target location of the transition chosen by the scheduler; and similarly, for demonic locations by replacing $\sigma$ by $\pi$ we
have
\[
Y_D:=\sum_{\loc\in L_D}\mathbf{1}_{\theta_n=\loc}\cdot\eta(\pi(\rho),\{\overline{x}_{k,n}\}_k),
\]
where $\rho$ is the finite path up to $n$ steps.
Then from properties of conditional expectation~\cite[Page 88]{probabilitycambridge}, one obtains:
\[
\expv^{\sigma,\pi}(X_{n+1}\mid\mathcal{F}_n)=\mathbf{1}_{\theta_n=\loc_p^\lout}\cdot X_n+Y'_P+Y'_S+Y_A+Y_D,
\]
because $\mathbf{1}_{\theta_n=\loc_p^\lout}\cdot X_n$, $Y_A$,  $Y_D$ are measurable in $\mathcal{F}_n$,
we have $\expv^{\sigma,\pi}(\mathbf{1}_{\theta_n=\loc_p^\lout}\cdot X_n \mid  \mathcal{F}_n ) =
\mathbf{1}_{\theta_n=\loc_p^\lout}\cdot X_n$ (and similarly for $Y_A$ and $Y_D$); and for $Y_P$ and $Y_S$
we need their expectation as $Y_P'$ and $Y_S'$ defined below.
We have
\[
Y'_P:=\sum_{\loc\in L_P}\Big[\mathbf{1}_{\theta_n=\loc}\cdot\sum_{i\in \{0,1\}} \big(\probm(B_\loc=i)\cdot\eta(\loc_{B_\loc=i},\{\overline{x}_{k,n}\}_{k})
\big)\Big]
\]
\[
 Y'_S:= \sum_{\loc\in L_S}\sum_{(\loc,f,\loc')\in\mapsto_\loc}\Big[\mathbf{1}_{\theta_n=\loc\wedge \{\overline{x}_{k,n}\}_{k}\models G(\tran)}\cdot \eta(\loc',\expv^{\sigma,\tau}_R\left(f(\{\overline{x}_{k,n}\}_{k},\mathbf{r})\right)\Big]\enskip.
\]
Note that when $\theta_n\in L_S \cup L_P \cup L_A$, by definition we have
$\mathrm{pre}_\eta(\theta_n, \{\overline{x}_{k,n}\}_{k}) = \mathbf{1}_{\theta_n=\loc_p^\lout}\cdot X_n+Y'_P+Y'_S+Y_A$;
and when $\theta_n \in L_D$ by definition we have
$Y_D\le \mathrm{pre}_\eta(\theta_n, \{\overline{x}_{k,n}\}_{k})$.
Hence the result follows.
\end{proof}

\noindent\textbf{Theorem~\ref{thm:supermartingale-correctness}}.
If there exists an LRSM $\eta$ wrt $I$ for $\mathcal{G}_P$, then
\begin{compactenum}
\item $P$ is a.s. terminating; and

\item the expected termination time of $P$ is at most
$\frac{\eta(\initloc,\initval)-K'}{\epsilon}$ and hence finite,
i.e., $\Eval(P) < \infty$.
\end{compactenum}

\begin{proof}
We establish both the points.
Let $\eta$ be an LRSM wrt $I$ for $\mathcal{G}_P$.
Define the angelic strategy $\sigma$ which solely depends on the end configuration of a finite path as follows:
\[
\sigma(\loc,\mathbf{x})=\mathrm{argmin}_{(\loc,f,\loc')\in\transitions}\eta(\loc',\mathbf{x})
\]
for all $\loc\in L_A$ and $\vec{x}\in\mathbb{R}^{|\pvars|}$.
Fix any demonic strategy $\pi$.
Let the stochastic process $\{X_n\}_{n\in\mathbb{N}}$ be defined by:
\[
X_n:=\eta(\theta_n,\{\overline{x}_{k,n}\}_k).
\]
For all $n\in\mathbb{N}$, from Lemma~\ref{lemma:condexp}, we have
\[
\expv^{\sigma,\pi}(X_{n+1}\mid\mathcal{F}_n)\le\mathrm{pre}_\eta(\theta_n,\{\overline{x}_{k,n}\}_{k})\enskip.
\]
By condition C4,
\[
\mathrm{pre}_\eta(\theta_n,\{\overline{x}_{k,n}\}_{k})\le \eta(\theta_n,\{\overline{x}_{k,n}\}_{k})-\epsilon\cdot \mathbf{1}_{\theta_n\ne\loc_P^\lout}
\]
for some $\epsilon\ge 1$.
Moreover, from C2, C3 and the fact that $I$ is a linear invariant, it holds almost-surely that $\theta_n\ne\loc_P^\lout$ iff $X_n\ge 0$.
Thus, we have
\[
\expv^{\sigma,\pi}(X_{n+1}\mid\mathcal{F}_n)\le X_n-\epsilon\cdot\mathbf{1}_{X_n\ge 0}\enskip.
\]
It follows that $\{X_n\}_{n\in\mathbb{N}}$ is an RSM.
Hence by Proposition~\ref{prop:rsm}, it follows that $\mathcal{G}_P$ terminates almost surely and
\[
\expv^{\sigma,\pi}(T)\le \frac{\eta(\initloc,\initval)-K'}{\epsilon}\enskip.
\]
The desired result follows.
\end{proof}

\section{Proofs for Section~\ref{subsec:algo}}

\noindent\textbf{Lemma~\ref{lemm:closedness}}.
Let $\mathbf{A}\in\mathbb{R}^{m\times n}$, $\mathbf{B}\in\mathbb{R}^{k\times n}$, $\mathbf{b}\in\mathbb{R}^{m}$, $\mathbf{d}\in\mathbb{R}^k$, $\mathbf{c}\in\mathbb{R}^{n}$ and $d\in\mathbb{R}$.
Let
$\mathbf{Z}_{<} = \{\mathbf{x}\mid \mathbf{A}\mathbf{x}\le \mathbf{b}\wedge \mathbf{B}\mathbf{x}< \mathbf{d}\}$ and
$\mathbf{Z}_{\leq} =\{\mathbf{x}\mid \mathbf{A}\mathbf{x}\le \mathbf{b}\wedge \mathbf{B}\mathbf{x}\leq \mathbf{d}\}$.
Assume that $\mathbf{Z}_{<} \neq \emptyset$.
Then for all closed subsets $H\subseteq\mathbb{R}^{|\pvars|}$ we have that
$\mathbf{Z}_{<} \subseteq H$ implies $\mathbf{Z}_{\leq} \subseteq H$.
\begin{proof}
Let $\mathbf{z}$ be any point such that $\mathbf{A}\mathbf{z}\le\mathbf{b}$, $\mathbf{B}\mathbf{z}\le\mathbf{d}$ and $\mathbf{B}\mathbf{z}\not<\mathbf{d}$.
Let $\mathbf{y}$ be a point such that $\mathbf{A}\mathbf{y}\le\mathbf{b}$ and $\mathbf{B}\mathbf{y}<\mathbf{d}$.
Then for all $\lambda\in (0,1)$,
\[
\lambda\cdot\mathbf{y}+(1-\lambda)\cdot\mathbf{z}\in \{\mathbf{x}\mid \mathbf{A}\mathbf{x}\le\mathbf{b}\wedge\mathbf{B}\mathbf{x}<\mathbf{d}\}\enskip.
\]
By letting $\lambda\rightarrow 1$, we obtain that $\mathbf{z}\in H$ because $H$ is closed.
\end{proof}

\noindent\textbf{Lemma~\ref{lemm:linearemptiness}}.
Let $\mathbf{A}\in\mathbb{R}^{m\times n}$, $\mathbf{B}\in\mathbb{R}^{k\times n}$, $\mathbf{b}\in\mathbb{R}^{m}$ and $\mathbf{d}\in\mathbb{R}^k$.
Then all of the following three problems can be decided in polynomial time in the binary encoding of $\mathbf{A},\mathbf{B},\mathbf{b},\mathbf{d}$:
\begin{enumerate}\itemsep1pt \parskip0pt \parsep0pt
\item $\{\mathbf{x}\mid \mathbf{A}\mathbf{x}\le \mathbf{b}\}\stackrel{?}{=}\emptyset$;
\item $\{\mathbf{x}\mid \mathbf{A}\mathbf{x}\le \mathbf{b}\wedge \mathbf{B}\mathbf{x}< \mathbf{d}\}\stackrel{?}{=}\emptyset$;
\item $\{\mathbf{x}\mid \mathbf{B}\mathbf{x}< \mathbf{d}\}\stackrel{?}{=}\emptyset$.
\end{enumerate}
\begin{proof}
The polynomial-time decidability of the first problem is well-known (cf. ~\cite{DBLP:books/daglib/0090562}).
The second problem can be solved by checking whether the optimal value of the following linear program
is greater than zero:
\[
\max z \text{ subject to }
\mathbf{A}\mathbf{x}\le\mathbf{b}; \quad \mathbf{B}\mathbf{x}+z\cdot\mathbf{1}\le\mathbf{d}; \quad z\ge 0.
\]
The proof for the third problem is similar to the second one.
\end{proof}

\section{Proofs for Section~\ref{subsec:hard1}}

\smallskip\noindent{\bf Lemma~\ref{lemm:hard1}.}
The \LRAPP\ realizability problem for \APP s with angelic non-determinism is 
$\NP$-hard, even for non-probabilistic non-demonic programs with simple guards.

\begin{proof}
We show a polynomial reduction from $3$-\textsc{SAT} to the \LRAPP\ realizability problem.

Let $\psi$ be any propositional formula in a conjunctive normal form with three literals per clause. Let $C_1,\dots,C_m$ be all the clauses and $x_1,\dots,x_n$ all the variables of $\psi$. For every $1 \leq j \leq m$ we write $C_j \equiv \lit_{j,1} \vee \lit_{j,2} \vee \lit_{j,3}$, where each $\lit_{j,k}$ is either a positive or a negative literal, i.e. a variable or its negation. 

We construct a program $P_{\psi}$ as follows: the program variables of $P_\psi$ correspond to the variables in $\psi$, the program has no random variables. All the program variables are initially set to 1. To construct the body of the program, we define, for each literal $\lit$ of $\psi$ involving a variable $x_i$, a linear expression $g_{\lit}$ which is equal either to $x_i$ or $1-x_i$ depending on whether $\lit$ is a positive literal or not, respectively. The body of the program then has the form 
\[
\mathbf{while}~\varphi~\mathbf{do}~Q_1;\cdots;Q_n~\mathbf{od} 
\]
where
\begin{itemize}
\item $\varphi$ is a finite disjunction of linear constraints of the form 
\[
\bigvee_{j=1}^{m} \left[g_{\lit_{j,1}} + g_{\lit_{j,2}} + g_{\lit_{j,3}}\leq \frac{1}{2}\right]\enskip;
\] 
\item $Q_i$ has the form \textbf{if angel then} $x_i:=1$ \textbf{else} $x_i:=0$ \textbf{fi}, for all $1\leq i \leq n$.
\end{itemize}

Clearly, given a formula $\psi$ the program $P_{\psi}$ can be constructed in time polynomial in the size of $\psi$, and moreover, $P_{\psi}$ is non-probabilistic and non-demonic.

Note that each valuation $\vec{x}$, reachable from the initial configuration of $\mathcal{G}_{P_{\psi}}$, can be viewed as a bit vector, and hence we can identify these reachable valuations with truth assignments to $\psi$. Also note that for such a reachable valuation $\vec{x}$ it holds $g_{\lit_{j,k}}(\vec{x})=1$ if and only if $\vec{x}$, viewed as an assignment, satisfies the formula $\lit_{j,k}$, and $g_{\lit_{j,k}}(\vec{x})=0$ otherwise.

Finally, let $I$ be an invariant assigning to every location a linear assertion $\bigwedge_{i=1}^n [0 \leq x_i \wedge x_i \leq 1]$. Note that the invariant $I$ is very simple and it is plausible that it can be easily discovered without employing any significant insights into the structure of $P_{\psi}$.

We claim that $P_{\psi}$ admits a linear ranking supermartingale with respect to $I$ if and only if there exists a satisfying assignment for $\psi$. First let us assume that $\psi$ is not satisfiable. Then, as noted above, for all reachable valuations $\vec{x}$
there is $1\leq j \leq m$ such that for all $\lit_{j,k}$, $1\leq k \leq 3$, it holds $g_{\lit_{j,k}}(\vec{x})=0$. It follows that $\varphi$ holds in every reachable valuation and hence the program never terminates. From Theorem~\ref{thm:supermartingale-correctness} it follows that $P_{\psi}$ does not admit an LRSM for any invariant $I$.

Let us now assume that there exists an assignment 
$\nu\colon \{x_1,\dots,x_n\} \rightarrow\{0,1\}$ 
satisfying $\psi$. We use $\nu$ to construct an LRSM for $P_{\psi}$ with respect to $I$. First, let us fix the following notation of locations of $\mathcal{G}_{P_{\psi}}$: by $\ell_i$ we denote the initial location of the sub-program $Q_i$, by $\ell_i^1$ and $\ell_i^0$ the locations corresponding to "then" and "else" branches of $Q_i$, respectively, and by $\ell_{n+1}$ and $\ell^{\lout}$ the initial and terminal locations of $\mathcal{G}_{P_{\psi}}$, respectively. 
Next, we fix a \emph{penalty} constant $\satpenalty = 4n+3$ and for every pair of indexes $1\leq i \leq n+1$, $1\leq j \leq n$ we define a linear expression $$h_{i,j} = \begin{dcases} 1-x_j & \text{if }\nu(x_j)=1\text{ and } i\leq j,\\
(1-x_j)\cdot \satpenalty & \text{if }\nu(x_j)=1\text{ and } i>j, \\
x_j & \text{if }\nu(x_j)=0\text{ and } i \leq j, \\
x_j\cdot\satpenalty & \text{if }\nu(x_j)=0\text{ and } i>j.
  \end{dcases}$$
Note that for $\vec{x}\in [0,1]^n$ the value $h_{i,j}(\vec{x})$ equals either $|\nu(x_j)-\vec{x}[j]|$ or $|\nu(x_j)-\vec{x}[j]|\cdot \satpenalty$ depending on whether $i\leq j$ or not.

Finally, we  define an LRSM $\mart$ as follows: for every $1 \leq i \leq n+1$ we put \begin{align*}\mart(\ell_i,\vec{x}) &= \sum_{j=1}^{n} h_{i,j}(\vec{x}) + (n-i+1)\cdot 2.
\end{align*}
Next, for every $1\leq i \leq n$ we put 
\begin{align*} 
\mart(\ell_{i}^{\nu(x_{i})},\vec{x}) &= \sum_{j=1}^{n} h_{i,j}(\vec{x}) + (2(n-i)+1)\\
\mart(\ell_{i}^{1-\nu(x_{i})},\vec{x}) &= n\cdot\satpenalty+2n+1.
\end{align*}
Finally, we put $\mart(\ell^{\lout},\vec{x})=-1/2$.

We show that $\mart$ is an LRSM wrt $I$.

As $h_{i,j}(\vec{x})\geq 0$ whenever $\vec{x}\in [0,1]^n$, it is easy to check that $\vec{x}\in \bigcup I(\loc)\Rightarrow\mart(\loc,\vec{x})\geq 0$ for every non-terminal location $\loc$. Next, for each $\vec{x}\in\Rset^n$ and each angelic state $\ell_i$ it holds that $\mart(\ell_i,\vec{x}) = \mart(\ell_i^{\nu(x_i)},\vec{x}) + 1$; hence $\mathrm{pre}_\mart(\ell_i,\vec{x})\leq\mart(\ell_i,\vec{x})-1$. Further, let us observe that for all $\vec{x}\in[0,1]^n$ and all $1\leq i \leq n+1$ we have $\mart(\ell_i,\vec{x})\leq n\cdot\satpenalty +2n$; thus for all $1\leq i\leq n$ it holds that  $\mathrm{pre}_\mart(\ell_i^{1-\nu(x_i)},\vec{x})\leq\mart(\ell_i^{1-\nu(x_i)},\vec{x})-1$. 
Also, for all such $\vec{x}$ and $i$ it holds that 
\begin{eqnarray*}
& &\mart(\ell_i^{\nu(x_i)},\vec{x}) \\
&=&\mart(\ell_{i+1},\assgn{x_i}{\nu(x_i)}(\vec{x})) + h_{i,i}(\vec{x})\\
& &\quad{}-h_{i,i+1}(\assgn{x_i}{\nu(x_i)}(\vec{x}))+ 1 \\
&\geq & \mart(\ell_{i+1},\assgn{x_i}{\nu(x_i)}(\vec{x})) + 1\enskip,
\end{eqnarray*}
where the latter inequality follows from $h_{i,i+1}(\assgn{x_i}{\nu(x_i)}(\vec{x}))=0$ and non-negativity of $h_{i,i}(\vec{x})$ for $\vec{x}\in[0,1]^n$. 
Hence, $\mathrm{pre}_\mart(\ell_i^{\nu(x_i)},\vec{x})\leq\mart(\ell_i^{\nu(x_i)},\vec{x})-1$. 
Finally, let us focus on $\ell_{n+1}$. Let $\vec{x}$ be such that $\vec{x}\models\varphi\wedge I(\loc_{n+1})$. 
It is easy to check that 
\begin{equation}
\label{eq:mart-diff-sat}
\mathrm{pre}_\mart(\ell_{n+1},\vec{x})=\mart(\ell_{n+1},\vec{x})+2n - (\satpenalty-1)\cdot\sum_{i=1}^{n} |\nu(x_i)-\vec{x}[i]|. \end{equation}
 Now since $\vec{x}\models \varphi$, there is $1\leq j \leq m$ such that $(g_{\lit_{j,1}} + g_{\lit_{j,2}} + g_{\lit_{j,3}})(\vec{x}) \leq  \frac{1}{2}$, which implies $g_{\lit_{j,k}}(\vec{x})\leq\frac{1}{2}$ for all $1\leq k\leq 3$, as all these numbers are non-negative. On the other hand, there is $1\leq k'\leq 3$ such that $\nu(\lit_{j,k'})=1$, as $\nu$ is a satisfying assignment. Let $x_i$ be the variable in $\lit_{j,k'}$. If $\lit_{j,k'}$ is a positive literal, then $\nu(x_i)=1$ and $\vec{x}[i]=g_{\lit_{j,k'}}(\vec{x}) \leq\frac{1}{2}$, and if $\lit_{j,k'}$ is negative, then $\nu_{x_i}=0$ and $\vec{x}[i]=1-g_{\lit_{j,k'}}(\vec{x}) \geq\frac{1}{2}$. In both cases we get $|\nu(x_i)-\vec{x}[i]|\geq\frac{1}{2}$. Plugging this into~\eqref{eq:mart-diff-sat} and combining with the fact that $(\satpenalty -1)/2\geq 2n+1$ we get that $\mathrm{pre}_\mart(\ell_{n+1},\vec{x})\leq\mart(\ell_{n+1},\vec{x})-1$. Hence, $\mart$ is indeed an LRSM.
\end{proof}

\section{Details of Section~\ref{sec:quan}}


\subsection{Irrationality of expected termination time}

Let us again consider the program $P$ in Figure~\ref{fig:irrational-time}. We claim that $\Eval(P)$ is irrational. Note that the stochastic game structure $\mathcal{G}_P$ has six locations, where the following two are of interest for us: the location corresponding to the beginning of the while loop, which we denote by $\textit{while}$, and the location entered after executing the first assignment in the else branch, which we denote by $\textit{else}$. 
Consider the following systems of polynomial equations:

\newcommand{\stalti}{\textit{if}}
\newcommand{\staltii}{\textit{else}}
\newcommand{\stalt}{\textit{while}}
\begin{align}
\nonumber x_{\stalt,\stalt} &= \frac{1}{2}(x_{\stalt,\stalt}^2 +  x_{\stalt,\staltii}) \\
\nonumber x_{\stalt,\staltii} &= \frac{1}{2}+\frac{1}{2}(x_{\stalt,\stalt}\cdot x_{\stalt,\staltii} ) \\
\nonumber y_{\stalt,\stalt} & = \frac{\frac{1}{2}(3 + 2y_{\stalt,\stalt}\cdot x^2_{\stalt,\stalt} + ( y_{\stalt,\staltii} + 1)\cdot x_{\stalt,\staltii})}{x_{\stalt,\stalt}}\\
\nonumber y_{\stalt,\staltii} & = \frac{\frac{3}{2}+\frac{1}{2}(3 + y_{\stalt,\stalt}+y_{\stalt,\staltii})\cdot x_{\stalt,\stalt}\cdot x_{\stalt,\staltii}}{x_{\stalt,\staltii}}\\
\label{eq:system}
\end{align}

Intuitively,  the variable $x_{\ell,\ell'}$ represents the probability that when starting in location $\ell$ with variable $n$ having some value $k$, the first location in which the value of $n$ decreases to $k-1$ is $\ell'$ (note that this probability is independent of $k$). Next, $y_{\ell,\ell'}$ is the conditional expected number of steps needed to decrease $n$ by 1, conditioned by the event that we start in $\ell$ and the first state in which $n$ drops below the initial value is $\ell'$. Formally, it is shown in~\cite{BEKK:pPDA-survey-FMSD} and~\cite{EKM:prob-PDA-expectations} that these probabilities and expectations are the minimal non-negative solutions of the system.

It is then easy to check that the expected termination time of $P$ is given by the formula
\begin{equation}
\label{eq:irrational-time}
1 + x_{\stalt,\stalt}(y_{\stalt,\stalt}+1) + x_{\stalt,\staltii}(y_{\stalt,\staltii}+2).\end{equation}
Examining the solutions of \eqref{eq:system} (we used \emph{Wolfram Alpha} to identify them) we get that $x_{\stalt,\stalt}=\frac{3-\sqrt{5}}{2}$, $x_{\stalt,\staltii}=\frac{\sqrt{5}-1}{2}$, $y_{\stalt,\stalt}=(85+43\sqrt{5})/10$, $y_{\stalt,\staltii}= (15+29\sqrt{5})/10$. Plugging this into~\eqref{eq:irrational-time} we get that $\Eval(P)=2(5+\sqrt{5})$.

\subsection{Proof of Lemma~\ref{coro:concen}}

%


\smallskip\noindent{\bf Lemma~\ref{coro:concen}.}
%
%
%
The quantitative approximation problem can be solved in a doubly exponential time for programs with only discrete probability choices.

\begin{proof}
The algorithm proceeds via analysing a suitable finite \emph{unfolding} of the input program $P$. Formally, an $n$-step unfolding of $P$, where $n\in \Nset$, is an SGS $\unfold{P}{n}$ such that
\begin{itemize}
\item the set locations of $\unfold{P}{n}$ consists of certain finite paths in $\mathcal{G}_P$ (i.e. in the SGS associated to $P$). Namely, we take those finite paths whose  starting configuration is the initial configuration of $\mathcal{G}_P$ and whose length is at most $n$. The type of such a location $w$ is determined by the type of a location in the last configuration of $\mathcal{G}_P$ appearing on $w$.
\item $\unfold{P}{n}$ has the same program and random variables as $\mathcal{G}_P$ (i.e. the same ones as $P$) and the initial configuration of $\unfold{P}{n}$ is the tuple $((\loc_0,\vec{x}_0),\vec{x}_0)$, where $(\loc_0,\vec{x}_0)$ is the initial configuration of $\mathcal{G}_P$ (it can be identified with a finite path of length 0).
\item For every pair $w$, $w'$ of locations of $\unfold{P}{n}$ (i.e., $w$, $w'$ are finite paths in $\mathcal{G}_P$) such that $w'$ can be formed by performing a single transition $\tau$ out of the last configuration of $w$, there is a transition $(w,f,w')$ in $\unfold{P}{n}$, where $f$ is an update function of $\tau$. Moreover, for every history $w$ of length $n$ that is also a location of $\unfold{P}{n}$ there is a transition $(w,\id,w)$ in $\unfold{P}{n}$.
\item The probability distributions and guard functions for probabilistic/deterministic locations $w $ of $\unfold{P}{n}$ are determined, in a natural way, by the probability distribution/guard functions of the location in the last configuration of $w$. 
\end{itemize}

Note that that each configuration of $\unfold{P}{n}$ reachable from the initial configuration is of the form $((\loc_0,\vec{x}_0),\dots,(\loc_i,\vec{x}_i) , \vec{x}_i)$, where $(\loc_j,\vec{x}_j), 0 \leq j \leq i$ are configurations of $\unfold{P}{n}$. In particular, the values of program variables in every reachable configuration $(w,\vec{x})$ of $\unfold{P}{n}$ are uniquely determined by $w$. To avoid confusion, in the following we will denote the locations in $\mathcal{G}_P$ by $\loc,\loc_0,\loc_1,\dots$ and locations in $\unfold{P}{n}$ by $w,w_0,w_1,\dots$.

Given $P$ and $n$ it is possible to construct $\unfold{P}{n}$ in time exponential in $n$ and in the encoding size of $P$. To see this, observe that the graph of $\unfold{P}{n}$ is a tree of depth $n$ and so $\unfold{P}{n}$ has at most $2^n$ locations. We can construct $\unfold{P}{n}$ by, e.g. depth-first enumeration of all finite paths in $\mathcal{G}_P$ of length $\leq n$. To bound the complexity of the construction we need to bound the magnitude of numbers (i.e. variable values) appearing on these finite paths -- these are needed to determine which transition to take in deterministic locations. Note that for every \textsc{App} $P$  there is a positive number $K$ at most exponential in the encoding size of $P$ such that in every step the absolute value of any variable of $P$ can increase by the factor of at most $K$ in every step. Hence, after $n$ steps the absolute value of every variable is bounded by $K^n\cdot L$, where $L$ is a maximal absolute value of any variable of $P$ upon initialization ($L$ is again at most exponential in the size of $P$). The exponential (in $n$ and size of $P$) bound on the time needed to construct $\unfold{P}{n}$ follows. 

After constructing $\unfold{P}{n}$, we post-process the SGS by removing all transitions outgoing from angelic locations that violate the supermartingale property of $\mart$. Formally, for every angelic location $w$ of $\unfold{P}{n}$ and each its successor $w'$ we denote by $c$ and $c'$ the last configurations on $w$ and $w'$, respectively. We then check whether $ \mathrm{pre}_\eta(c)= \mart(c')$, and if not, we remove the transition from $w$ to $w'$.

There is a natural many-to-one surjective correspondence $\Gamma$ between demonic schedulers in $\mathcal{G}_P$ and demonic schedulers in $\unfold{P}{n}$ (the schedulers in $\mathcal{G}_P$ that have the same behaviour up to step $n$ are mapped to a unique scheduler in $\unfold{P}{n}$ which induces this behaviour). Similarly, there is a many-to-one correspondence $\Delta$ between angelic schedulers that belong to $\martcor{\mart}$ in $\mathcal{G}_P$ and angelic schedulers in $\unfold{P}{n}$.

Let $\term_n$ be the set of all locations $w$ in $\unfold{{P}}{n}$ that, when interpreted as paths in $\mathcal{G}_P$, end with a configuration of the form $(\loc_P^{\lout},\vec{x})$ for some $\vec{x}$. It is easy to verify that for any adversarial scheduler $\pi$ in $\mathcal{G}_P$ the probability $\probm^{\pi}(T\leq x))$ is equal to the probability of reaching $\term_n$ under $\Phi(\pi)$ in $\unfold{P}{n}$. 

To solve the quantitative approximation problem we proceed as follows: First we compute number $n$, at most exponential in the size of $P$, such that the infimum (among all angelic schedulers in $\martcor{\mart}$) probability of  \emph{not terminating} in the first $n$ steps is "sufficiently small" in $\mathcal{G}_P$. Then we construct, in time exponential in $n$ and size of $P$ the unfolding $\unfold{P}{n}$ and assign a non-negative cost $\stepcost(w)$ to each location $w$ of $\unfold{P}{n}$. We then define a random variable $\costvar$ which to each run $(w_0,\vec{x}_0),(w_1,\vec{x}_1),\dots $ in $\unfold{P}{n}$ assigns the number $\sum_{i=0}^{j} \stepcost(w_i)$, where $j=\min\{n,\min\{k\mid w_k \in \term_n\}\}$. Using the fact that in $\mathcal{G}_P$ we terminate with very high probability in the first $n$ steps, we will show how to construct the cost function $\stepcost$ in such a way that for every pair of schedulers $\sigma,\pi$, where $\sigma\in \martcor{\mart}$, it holds $|\E^{\sigma,\pi}T-\E^{\Delta(\sigma),\Gamma(\pi)}\costvar|\leq \delta$ (the construction of $\stepcost$ can be done in time polynomial in the size of $\unfold{P}{n}$). Hence, to solve the quantitative approximation problem it will suffice to compute $\inf_{\sigma'}\sup_{\pi'}\E^{\sigma',\pi'}\costvar$, where the infimum and supremum are taken over schedulers in $\unfold{P}{n}$. This computation can be done in polynomial time via standard backward iteration~\cite{OR:book}. The doubly-exponential bound then follows from the aforementioned bound on the size of $\unfold{P}{n}$.

First let us show how to compute $n$. Since $P$ is a bounded \LRAPP and we are given the corresponding ranking supermartingale $\mart$ with one-step change bounds $a,b$, by Theorem~\ref{thm:quan_lrapp} we can compute, in polynomial time, a concentration bound $B$ which is at most exponential in the size of $P$. That is, there are positive numbers $ c_1$, $c_2$, which depend just on $P$, such that for each $x\geq B$ it holds $\Thr(P,x)\leq c_1\cdot \exp(-c_2\cdot x)$. Moreover, numbers $c_1$ and $c_2$ are also computable in polynomial time and at most exponential in the size of $P$, as witnessed in the proof of Theorem~\ref{thm:quan_lrapp}. Now denote by $M_0$ the value of the LRSM $\mart$ in the initial configuration of $P$. Note that $M_0$, as well as the bounds $a$, $b$, are at most exponential in the size of $P$ as $\mart$, $a$ and $b$ can be obtained by solving a system of linear inequalities with coefficients determined by $P$. We compute (in polynomial time) the smallest integer $n\geq B$ such that $ c_1 \cdot e^{-c_2 \cdot n} \cdot (n + (M_0+n\cdot (b-a))/\eps) \leq \delta$, where $\eps$ is as in Definition~\ref{def:lrsm} (the minimal expected decrease of the value of $\mart$ in a single step). It is straightforward to verify that such an $n$ exists and that it is polynomial in $M_0$, $\eps$, and $a,b$, and hence exponential in the size of $P$.

Now let us construct the cost function $\stepcost$. The initial location gets cost 0 and all other locations get cost 1 \emph{except} for locations $w$ that represents the paths in $\mathcal{G}_P$ of lenth $n$. The latter locations get a special cost $C = (M_0 + n\cdot (b-a)  )/\eps$.

We prove that for any pair of schedulers $\sigma,\pi$, $\sigma\in \martcor{\mart}$, it holds $|\E^{\sigma,\pi}T-\E^{\Delta(\sigma),\Gamma(\pi)}\costvar|\leq \delta$. So fix an arbitrary such pair. 
We start by proving that $\E^{\sigma,\pi}T \leq \E^{\Delta(\sigma),\Gamma(\pi)}\costvar$. Since the behaviour of the schedulers in the first $n$ steps is mimicked by $\Delta(\sigma)$ and $\Gamma(\pi)$ in $\unfold{P}{n}$, and in the latter SGS we accumulate one unit of cost per each of the first $n-1$ steps, it suffices to show that the expected number of steps needed to terminate with these schedulers from any configuration of $\mathcal{G}_P$ that is reachable in exactly $n$ steps is bounded by $C$. Since the value of $\mart$ can change by at most $(b-a)$ in every step, and $\sigma\in\martcor{\mart}$, the value of $\mart$ after $n$ steps in $\mathcal{G}_P$ is at most $M_0 + n\cdot(b-a)$. By Theorem~\ref{thm:supermartingale-correctness} the expected number of step to terminate from a configuration with such an $\mart$-value is at most $(M_0 + n\cdot(b-a))/\eps$ as required.

To finish the proof we show that $\E^{\Delta(\sigma),\Gamma(\pi)}\costvar \leq \E^{\sigma,\pi}T + \delta$. First note that $\E^{\Delta(\sigma),\Gamma(\pi)}\costvar - \E^{\sigma,\pi}T$ can be bounded by $p\cdot C$, where $p$ is the probability that a run in $\mathcal{G}_P$ does not terminate in first $n$ steps under these schedulers. Since $n\geq B$ it holds $p \leq c_1 \cdot e^{-c_2 \cdot n}$. Thus $p\cdot C \leq c_1 \cdot e^{-c_2 \cdot n} \cdot (n + (M_0+n\cdot (b-a))/\eps) \leq \delta$, the last inequality following from the choice of $n$. 

\end{proof}

\subsection{Proofs for Section~\ref{subsec:hard2}}

\smallskip\noindent{\bf Lemma~\ref{lemm:hard2}.}
For every $C\in\Nset$ the following problem is $\PSPACE$-hard:
Given a program $P$ without probability or non-determinism, with simple guards, and belonging to bounded \LRAPP;
and a number $N \in \Nset$ such that either
$\Eval(P)\leq N$ or $\Eval(P)\geq N\cdot C$, decide, which of these two alternatives hold.

\begin{proof}[Proof]
Let us fix a constant $C$.

We start by noting that there exists a constant $K$ such that the following $K$-\textsc{linearly bounded membership problem} is $\PSPACE$-hard: 
Given a deterministic Turing machine (DTM) $\mathcal{T}$ such that on every input of length $n$ the machine $\mathcal{T}$ uses at most $K\cdot n$ tape cells, and given a word $w$ over the input alphabet of $\mathcal{T}$, decide, whether $\mathcal{T}$ accepts $w$. This is because there exists $K$ for which there is a DTM $\mathcal{T}_{\textsc{QBF}}$ satisfying the above condition which decides the \textsc{QBF} problem ($\mathcal{T}_\textsc{QBF}$ works by performing a simple recursive search of the syntax tree of the input formula). We show a polynomial-time reduction from this membership problem to our problems.

Let $\mathcal{T},w$ be an instance of the membership problem. Since $\mathcal{T}$ has a linearly bounded complexity with a known coefficient $K$, there is a number $J$ computable in time polynomial in size of $\mathcal{T}$ and $w$ such that if $\mathcal{T}$ accepts $w$, it does so in at most $J$ steps (note that the magnitude of $J$ is exponential in size of $\mathcal{T}$ and $w$). The intuition behind the reduction is as follows: we construct a deterministic affine program $P$ simulating the computation of $\mathcal{T}$ on input $w$. The program consists of a single while-loop guarded by an expression $m\geq 1 \wedge r \geq 1$, where $m$ is a "master" variable of the program initialized in the preamble to $C\cdot J\cdot |w|$, while $r$ is initialized to $1$. The body of the while loop encodes the transition function of $\mathcal{T}$. The current configuration of $\mathcal{T}$ is encoded in variables of $P$: for every state $s$ we have a variable $x_s$ which is equal to $1$ when the current state is $s$ and $0$ otherwise; next, for every $1\leq i \leq K\cdot |w|$, where $|w|$ is the length of $w$, and every symbol $a$ of the tape alphabet of $\mathcal{T}$ we have a variable $x_{i,a}$, which is equal to $1$ if $a$ is currently on the $i$-th tape cell, and 0 otherwise; and finally we have a variable $x_{\mathit{head}}$ which stores the current position of the head. Additionally, we add a variable $x_{\mathit{step}}$ which records, whether the current configuration was already updated during the current iteration of the while-loop. The variables are initialized so as to represent the initial configuration of $\mathcal{T}$ on input $w$, e.g. $x_{\mathit{head}}=1$ and $x_{i,a}=1$ if and only if either $i\leq |w|$ and $a$ is the $i$-th symbol of $w$, or $i>|w|$ and $a$ is the symbol of an empty cell. It is then straightforward to encode the transition function using just assignments and if-then-else statements, see below. At the end of each iteration of the while loop the master variable $m$ is decreased by 1. However, if the current state is also an accepting state of $\mathcal{T}$, then $r$ is immediately set to $0$ and thus the program terminates.  

More formally, to a transition $\delta$ of $\mathcal{T}$ saying that in a configuration $(s,a)$ the state should be changed to $s'$, the symbol rewritten by $b$, and the head moved by $h\in \{-1,0,+1\}$, we assign the following affine program $Q_{\delta}$ in which a guard $\varphi_{\delta} = \bigvee_{i=1}^{K\cdot |w|} (x_s = 1$ \textbf{and} $x_{\mathit{step}}=0$ \textbf{and} $x_{\mathit{head}}=i$ \textbf{and} $x_{i,a}=1)$ is used:

\lstset{language=affprob}
\lstset{tabsize=3}
\newsavebox{\pspacegadget}
\begin{lrbox}{\pspacegadget}
\begin{lstlisting}[mathescape]
if $\varphi_\delta$ then 
	$x_{s} := 0$; $x_{s'}:=1$; $x_{i,a}:=0$; $x_{i,b}:=1$; 
	$x_{\mathit{head}}:=x_{\mathit{head}}+h$; $x_{\mathit{step}}=1$
else
	skip
fi
if $x_{s_{\mathit{acc}}} = 1$ then $r:=0$ else skip fi;
\end{lstlisting}
\end{lrbox}
\begin{figure}[h]
\centering
\usebox{\pspacegadget}
\end{figure}

The overall program then looks as follows:

\newsavebox{\pspacered}
\begin{lrbox}{\pspacered}
\begin{lstlisting}[mathescape]
while $m\geq 1 \wedge r \geq 1$ do
	$m:=m-1$;
	$x_{\mathit{step}}:=0$
	$Q_{\delta_1}$; $\cdots$ $Q_{\delta_m}$
od,
\end{lstlisting}
\end{lrbox}
\begin{figure}[h]
\centering
\usebox{\pspacered}
\end{figure}
where $\delta_1,\dots,\delta_m$ are all transitions of $\mathcal{T}$. Clearly the program always terminates and belongs to bounded \LRAPP: there is a trivial bounded LRSM whose value in the beginning of the while loop is $m$ while in further locations inside the while loop it takes the form $m+d$, for suitably small constants $d$. Since $m$ changes by at most 1 in every step, this LRSM is bounded.

Each iteration of the while loop takes the same amount of steps, since exactly one of the programs $Q_{\delta_j}$ enters the if-branch (we can assume that the transition function of $\mathcal{T}$ is total). Let us denote this number by $W$. 
It is easy to see that if $\mathcal{T}$ \emph{does not} accept $w$, then the program terminates in exactly $C\cdot J\cdot W$ steps. On the other hand, if $\mathcal{T}$ does accept $w$, then the program terminates in at most $J\cdot W$ steps. To finish the reduction, we put the number $N$ mentioned in item 1. equal to $ J\cdot W$.  

\end{proof}


\section{Proofs for Section~\ref{subsec:concentration}}

\noindent\textbf{Theorem~\ref{thm:hoeffding}}.
Let $\{X_n\}_{n\in\mathbb{N}}$ be a supermartingale wrt some filtration $\{\mathcal{F}_n\}_{n\in\mathbb{N}}$ and $\{[a_n,b_n]\}_{n\in\mathbb{N}}$ be a sequence of intervals of positive length in $\mathbb{R}$.
If $X_1$ is a constant random variable and $X_{n+1}-X_n\in [a_n,b_n]$ a.s. for all $n\in\mathbb{N}$, then
\[
\mathbb{P}(X_n-X_1\ge\lambda)\le e^{-\frac{2\lambda^2}{\sum_{k=2}^n(b_k-a_k)^2}}
\]
for all $n\in\mathbb{N}$ and $\lambda> 0$.
\begin{proof}
The proof goes through the characteristic method similar to the original proof of Hoeffding's Inequality~\cite{Hoeffding1963inequality,ColinMcDiarmid1998concentration},
and hence we present only the important details.
For each $n\ge 2$ and $t> 0$, we have (using standard arguments similar to~\cite{Hoeffding1963inequality,ColinMcDiarmid1998concentration}):
\begin{eqnarray*}
\mathbb{E}(e^{tX_n}) &=& \mathbb{E}\left(\mathbb{E}(e^{tX_n}\mid\mathcal{F}_{n-1})\right)\\
&=& \mathbb{E}\left(\mathbb{E}\left(e^{t(X_n-X_{n-1})}\cdot e^{tX_{n-1}}\mid\mathcal{F}_{n-1}\right)\right)\\
&=& \mathbb{E}\left(e^{tX_{n-1}}\right)\cdot\mathbb{E}\left(\mathbb{E}\left(e^{t(X_n-X_{n-1})}\mid\mathcal{F}_{n-1}\right)\right)\enskip.
\end{eqnarray*}
Note that for all $x\in [a_n,b_n]$,
\begin{eqnarray*}
e^{tx} &\le & \frac{x-a_n}{b_n-a_n}\cdot e^{ta_n}+\frac{b_n-x}{b_n-a_n}\cdot e^{tb_n} \\
       &\le & \frac{x}{b_n-a_n}\cdot (e^{tb_n}-e^{ta_n})+\frac{b_ne^{tb_n}-a_ne^{ta_n}}{b_n-a_n}\enskip.\\
\end{eqnarray*}
Then from $X_n-X_{n-1}\in [a_n,b_n]$ a.s. and $\mathbb{E}(X_n\mid \mathcal{F}_{n-1})\le X_{n-1}$, we obtain
\[
\mathbb{E}(e^{tX_n})\le \mathbb{E}\left(e^{tX_{n-1}}\right)\cdot\frac{b_ne^{tb_n}-a_ne^{ta_n}}{b_n-a_n}\enskip.
\]
By applying the analysis in the proof of~\cite[Lemma 2.6]{ColinMcDiarmid1998concentration} (taking $a=a_n$ and $b=b_n$), we obtain
\[
\frac{b_ne^{tb_n}-a_ne^{ta_n}}{b_n-a_n}\le e^{\frac{1}{8}t^2(b_n-a_n)^2}\enskip.
\]
This implies that
\[
\mathbb{E}(e^{tX_n})\le \mathbb{E}\left(e^{tX_{n-1}}\right)\cdot e^{\frac{1}{8}t^2(b_n-a_n)^2}\enskip.
\]
By induction, it follows that
\begin{eqnarray*}
\mathbb{E}(e^{tX_n}) &\le & \mathbb{E}\left(e^{tX_{1}}\right)\cdot e^{\frac{1}{8}t^2\sum_{k=2}^n(b_k-a_k)^2} \\
& = & e^{t\mathbb{E}(X_{1})}\cdot e^{\frac{1}{8}t^2\sum_{k=2}^n(b_k-a_k)^2}\enskip.
\end{eqnarray*}
Thus, by Markov Inequality, for all $\lambda>0$,
\begin{eqnarray*}
\mathbb{P}(X_n-X_1\ge\lambda) 
&=& \mathbb{P}(e^{t(X_n-X_1)}\ge e^{t\lambda}) \\
&\le & e^{-t\lambda}\cdot \mathbb{E}\left(e^{t(X_n-X_1)}\right) \\
&\le & e^{-t\lambda+\frac{1}{8}t^2\sum_{k=2}^n(b_k-a_k)^2}\enskip.
\end{eqnarray*}
By choosing $t=\frac{4\lambda}{\sum_{k=2}^n(b_k-a_k)^2}$, we obtain 
\[
\mathbb{P}(X_n-X_1\ge\lambda)\le e^{-\frac{2\lambda^2}{\sum_{k=2}^n(b_k-a_k)^2}}\enskip.
\]
The desired result follows.
\end{proof}

\noindent\textbf{Proposition~\ref{prop:hoeffding}}.
$\{Y_n\}_{n\in\mathbb{N}}$ is a supermartingale and $Y_{n+1}-Y_n\in [a+\epsilon,b+\epsilon]$ almost surely for all $n \in \Nats$.
\begin{proof}
Recall that
\[
Y_n=X_n+\epsilon\cdot(\min\{T,n\}-1)\enskip.
\]
Consider the following random variable:
\[
U_n=\min\{T, n+1\}-\min\{T,n\}\enskip,
\]
and observe that this is equal to $\mathbf{1}_{T>n}$.
From the properties of conditional expectation~\cite[Page 88]{probabilitycambridge} and the facts
that (i) the event $T>n$ is measurable in $\mathcal{F}_n$
(which implies that $\expv(\mathbf{1}_{T>n}\mid \mathcal{F}_n)= \mathbf{1}_{T>n}$);
and (ii) $X_n\ge 0$ iff $T>n$ (cf. conditions C2 and C3),
we have
\begin{eqnarray*}
\expv(Y_{n+1}\mid\mathcal{F}_n)-Y_n
&=&\expv(X_{n+1}\mid\mathcal{F}_n)-X_n 
+\epsilon\cdot\expv(U_n \mid\mathcal{F}_n) \\
&=&\expv(X_{n+1}\mid\mathcal{F}_n)-X_n+\epsilon\cdot\expv(\mathbf{1}_{T>n}\mid\mathcal{F}_n)\\
&=&\expv(X_{n+1}\mid\mathcal{F}_n)-X_n+\epsilon\cdot\mathbf{1}_{T>n} \\
&\le &-\epsilon\cdot\mathbf{1}_{X_n\ge 0}+\epsilon\cdot\mathbf{1}_{T>n}\\
&=& 0\enskip.
\end{eqnarray*}
Note that the inequality above is due to the fact that $X_n$ is a ranking supermartingale.
Moreover, since $T\le n$ implies $\theta_n=\loc_P^\lout$ and $X_{n+1}=X_n$ we have that
$(X_{n+1}-X_n)= \mathbf{1}_{T>n}\cdot (X_{n+1}-X_n)$.
Hence we have
\begin{eqnarray*}
Y_{n+1}-Y_n &=& X_{n+1}-X_n+\epsilon\cdot U_n \\
&=& (X_{n+1}-X_n) +\epsilon\cdot \mathbf{1}_{T>n}   \\
&=& \mathbf{1}_{T>n}\cdot (X_{n+1}-X_n+\epsilon)\enskip.  \\
\end{eqnarray*}
Hence $Y_{n+1}-Y_n\in [a+\epsilon,b+\epsilon]$.
\end{proof}

\smallskip\noindent{\bf Applying Bernstein Inequality for incremental programs in \LRAPP.}
Let $\eta$ be an incremental LRSM template to be synthesized for $P$ and $\{X_n\}_{n\in\mathbb{N}}$ be the stochastic process defined by
\[
X_n:=\eta(\theta_n, \{\overline{x}_{k,n}\}_k)\enskip.
\]
To apply Bernstein's Inequality, we need to synthesize constants $c,M$ to fulfill that for all $n\in\Nset$, 
\[
\sqrt{\mathrm{Var}(X_{n+1}\mid \mathcal{F}_n)}\le c\mbox{ and } X_{n+1}-\expv(X_{n+1}\mid \mathcal{F}_n)\le M~.
\]
These conditions can be encoded by the following formulae:
\begin{itemize}
\item for all $\loc\in L_P$ with successor locations $\loc_1,\loc_2\in L$ and the branch probability value $p$, the sentence
\begin{align*}
&\forall\mathbf{x}.\forall i\in\{1,2\}.\Big(\eta(\loc_i,\mathbf{x})-\mathrm{pre}_\eta(\loc,\mathbf{x})\le M\\
&\wedge\sqrt{p(1-p)}\cdot\left|b^{\loc_1}-b^{\loc_2}\right|\le c \Big)
\end{align*}
holds, where one can obtain from easy calculation (on Bernoulli random variables) that
\[
\sqrt{\mathbf{1}_{\theta_n=\loc}\cdot\mathrm{Var}(X_{n+1}\mid\mathcal{F}_n)}\equiv\mathbf{1}_{\theta_n=\loc}\cdot\sqrt{p(1-p)}\cdot\left|b^{\loc_1}-b^{\loc_2}\right|\enskip;
\]
\item
for all $\loc\in L_S$ and all $\tran=(\loc,f,\loc')\in \mapsto_\loc$, the sentence
\begin{align*}
&\forall\mathbf{x}.\forall\vec{r}.\Big(\eta(\loc',f(\vec{x},\vec{r}))-\eta(\loc',\expv_R(f(\vec{x},\vec{r})))\le M\\
&\quad\wedge\mathbf{a}^{\loc}_x\cdot\sqrt{\mathrm{Var}_R(f(\vec{x},\vec{r}))}\le c\Big)
\end{align*}
holds, for which
\[
\sqrt{\mathbf{1}_{\theta_n=\loc}\cdot\mathrm{Var}(X_{n+1}\mid \mathcal{F}_n)}\equiv \mathbf{1}_{\theta_n=\loc}\cdot\mathbf{a}^{\loc}_x\cdot\sqrt{\mathrm{Var}_R(f(\vec{x},\vec{r}))}
\]
where $x$ is the updated program variable, $\mathbf{a}^{\loc}_x$ is the coefficient vector variable of $\vec{a}^\loc$ on program variable $x$ and $\mathrm{Var}_R(f(\vec{x},\vec{r}))$ is the variance on random variables $\vec{r}$.
\end{itemize}
Note that the formulae above can be transformed into existentially-quantified non-strict linear assertions by Farkas' Lemma.
Also note that there are no conditions for angelic or demonic locations since once the angelic
and demonic strategies are fixed, then there is no stochastic behavior in one step for such locations,
and hence the variance is zero.

Then we can apply Bernstein's Inequality in the same way as for Hoeffding's Inequality.
Define the stochastic process $\{Z_n\}_{n\in\mathbb{N}}$ by
\[
Z_n=X_n+\epsilon\cdot(\min\{T,n\}-1)\enskip;
\]
similar to $Y_n$ for Proposition~\ref{prop:hoeffding}.
We have the following proposition.

\begin{proposition}\label{prop:bernstein}
$\{Z_n\}_{n\in\mathbb{N}}$ is a supermartingale. Moreover, $Z_{n+1}-\expv(Z_{n+1}\mid \mathcal{F}_n)\le M$ and $\mathrm{Var}(Z_{n+1}\mid\mathcal{F}_n)\le c^2$,
for all $n \in \Nats$.
\end{proposition}
\begin{proof}
By the same analysis in Proposition~\ref{prop:hoeffding}, we have
$\{Z_n\}$ is a supermartingale.
Consider the following random variable:
$U_n=\min\{T, n+1\}-\min\{T,n\}$, which is equal to $\mathbf{1}_{T>n}$.
From the properties of conditional expectation~\cite[Page 88]{probabilitycambridge} and the facts
that (i) the event $T>n$ is measurable in $\mathcal{F}_n$
(which implies that $\mathrm{Var}(\mathbf{1}_{T>n}(X_{n+1}-X_n)\mid \mathcal{F}_n)= \mathbf{1}_{T>n}\mathrm{Var}(X_{n+1}-X_n\mid \mathcal{F}_n)$) we have the following:

\[
Z_{n+1}-\expv(Z_{n+1}\mid \mathcal{F}_n) \le X_{n+1}-\expv(X_{n+1}\mid\mathcal{F}_n) \le M
\]
and
\begin{eqnarray*}
\mathrm{Var}(Z_{n+1}\mid\mathcal{F}_n) &=& \mathrm{Var}(Z_{n+1}-Z_n\mid\mathcal{F}_n) \\
&=& \mathrm{Var}(X_{n+1}-X_n+ \epsilon \cdot U_n \mid \mathcal{F}_n) \\
&=& \mathrm{Var}(X_{n+1}-X_n+\epsilon \cdot \mathbf{1}_{T>n} \mid \mathcal{F}_n) \\
&\stackrel{*}{=}& \mathrm{Var}(\mathbf{1}_{T>n}\cdot (X_{n+1}-X_n+\epsilon)\mid \mathcal{F}_n)\\
&=& \mathbf{1}_{T>n}\cdot\mathrm{Var}(X_{n+1}-X_n\mid \mathcal{F}_n)\\
&\le & c^2\enskip.
\end{eqnarray*}
where (*) follows from the fact that $T\le n$ implies $\theta_n=\loc_P^\lout$ and $X_{n+1}=X_n$,
and the last inequality follows the fact that $ \mathrm{Var}(X_{n+1}-X_n\mid \mathcal{F}_n) \leq c^2$
since $c$ is obtained from the synthesis of the LRSM.
\end{proof}

Then similar to the derivation for Hoeffding's Inequality, we have
\begin{eqnarray*}
\mathbb{P}(T>n)
&\le & \mathbb{P}(Z_n-Z_1\ge \epsilon(n-1)-W_0)\\
&\le & e^{-\frac{(\epsilon(n-1)-W_0)^2}{2c^2(n-1)+\frac{2}{3}\cdot M(\epsilon(n-1)-W_0)}}\enskip.
\end{eqnarray*}
The optimal choice of concentration threshold is the same as for Hoeffding's Inequality, and
the optimality of upper bounds is reduced to a binary search on $z$ satisfying
\[
\frac{(\epsilon(n-1)-W_0)^2}{2c^2(n-1)+\frac{2}{3}\cdot M(\epsilon(n-1)-W_0)}\ge z
\]
and the constraints for LRSMs, the constraint $\epsilon\cdot (n-2)\ge W_0$ and the constraint for $c,M$,
which can be solved by quadratic programming~\cite{QuadraticProgrammingVavasis}.

\section{Details related to Experimental Results}\label{sec:app_exp}
We present below the details of the code modeling the various 
random walks of Section~\ref{sec:exp}, 
along with the invariants specified in the brackets.
Also in the description we use $\mathrm{Unif}$ to denote the uniform 
distribution.

\smallskip\noindent{\em Explanation why RW in 2D not bounded \LRAPP.}
We explain why the example RW in 2D is not a bounded \LRAPP: 
in this example at any point either the $x$ or the $y$ coordinates 
changes by at most~2, hence intuitively, the difference between 
two steps is bounded. 
However, to exploit such a fact one needs to consider a martingale
defining $\min\{x,y\}$, i.e., the minimum of two coordinate.
However such a martingale is not linear.
For this example, the LRSM is not a bounded martingale as the difference
between the two coordinates can be large.

\lstset{language=affprob}
\lstset{tabsize=3}
\newsavebox{\figonerwint}
\begin{lrbox}{\figonerwint}
\begin{lstlisting}[mathescape]

x:=n;    
[$x\ge -1$]                            
while $x\geq 0$ do
  [$x\ge 0$]
  if $\textbf{prob}\mathbf{(}0.3\mathbf{)}$ then
     [$x\ge 0$]
     $x:=x+1$
  else 
    [$x\ge 0$]
    $x:=x-1$; 
 fi 
od

[$x< 0$]
\end{lstlisting}
\end{lrbox}
\begin{figure}[h]
\centering
\usebox{\figonerwint}
\caption{Integer-valued random walk in one dimension, along with linear
invariants in square brackets.}
\label{fig:rwone}
\end{figure}

\lstset{language=affprob}
\lstset{tabsize=3}
\newsavebox{\figonerwuni}
\begin{lrbox}{\figonerwuni}
\begin{lstlisting}[mathescape]
$x:=n$;
[$x\ge -1$]
while $x\geq 0$ do
  [$x\ge 0$]
  if $\textbf{prob}\mathbf{(}0.3\mathbf{)}$ then 
     [$x\ge 0$]
     $x:=x + \mathrm{Unif}[0,1]$
  else 
     [$x\ge 0$]
     $x:=x - \mathrm{Unif}[0,1]$; 
  fi 
od

[$x<0$]
\end{lstlisting}
\end{lrbox}
\begin{figure}[h]
\centering
\usebox{\figonerwuni}
\caption{Real-valued random walk in one dimension, along with linear 
invariants in square brackets.}
\label{fig:rwtwo}
\end{figure}

\lstset{language=affprob}
\lstset{tabsize=3}
\newsavebox{\onedimadv}
\begin{lrbox}{\onedimadv}
\begin{lstlisting}[mathescape]
[$x\ge 0$]

while $n\geq 0$ do 
    [$x\ge 0$]
    $x:= x + r$;
    [$x\ge 0$]
    if demon then
        [$x\ge 0$] 
        if $\textbf{prob(}\frac{7}{8}\textbf{)}$ then 
          [$x\ge 0$]        
          $x:=x-1$ 
        else 
          [$x\ge 0$]        
          $x:= x+1$ 
        fi
    else
        [$x\ge 0$] skip; [$x\ge 0$] $x:=x-1$
    fi 
od

[$x< 0$]
\end{lstlisting}
\end{lrbox}
\begin{figure}[h]
\centering
\usebox{\onedimadv}
\caption{A queuing example: Adversarial random walk in one dimension.
The linear invariants in square brackets. The distribution of the random variable $r$ is described on page~\pageref{sec:experiments}}.
\label{fig:queue}
\end{figure}

\lstset{language=affprob}
\lstset{tabsize=3}
\newsavebox{\twodimadv}
\begin{lrbox}{\twodimadv}
\begin{lstlisting}[mathescape]

$x:=n_1$; $y=n_2$;

[$x\ge -2\wedge y\ge -3$]
 
while ($x\ge 0\wedge y\ge 0$) do
   [$x\ge  0\wedge y\ge 0$]
   if (demon) 
        [$x\ge  0\wedge y\ge 0$]
        $x:=x + \mathrm{Unif}[-2,1]$
   else 
        [$x\ge  0\wedge y\ge 0$]
        $y:=y + \mathrm{Unif}[-2,1]$
od
 
[$x<0 \vee y<0$]
\end{lstlisting}
\end{lrbox}
\begin{figure}[h]
\centering
\usebox{\twodimadv}
\caption{A 2D random walk example: adversarial random walk in two dimension, 
along with linear invariants in square brackets.}
\label{fig:2drandomwalk}
\end{figure}

\lstset{language=affprob}
\lstset{tabsize=3}
\newsavebox{\twodimadvvariant}
\begin{lrbox}{\twodimadvvariant}
\begin{lstlisting}[mathescape]

$x:=n_1$; $y=n_2$;

[$x-y\ge -3$]

while ($x\ge y$) do

[$x\ge y$]

   if (demon)  [$x\ge y$]
       if (prob(0.7)) 
          [$x\ge y$] $x:=x + \mathrm{Unif}[-2,1]$ 
       else 
          [$x\ge y$] $y:=y + \mathrm{Unif}[-2,1]$
   else [$x\ge y$]
       if (prob(0.7)) 
          [$x\ge y$] $y:=y + \mathrm{Unif}[2,-1]$ 
       else 
          [$x\ge y$] $x:=x + \mathrm{Unif}[2,-1]$
od

[$x<y$]
\end{lstlisting}
\end{lrbox}
\begin{figure}[h]
\centering
\usebox{\twodimadvvariant}
\caption{A 2D random walk example: Variant adversarial random walk in two dimension. 
The linear invariants in square brackets.}
\label{fig:2drandomwalkvariant}
\end{figure}

\end{document}